\documentclass[10pt,doublecolumn]{IEEEtran}
\usepackage{amsmath,amsthm,amsfonts,amssymb,bm,mathrsfs}
\usepackage{graphicx,subfigure}
\usepackage[cmintegrals]{newtxmath}
\usepackage{cite}
\usepackage{setspace}
\usepackage{caption}
\usepackage[usenames,dvipsnames]{color}
\usepackage{colortbl}
\definecolor{mygray}{gray}{.9}

\ifCLASSINFOpdf
\else
\fi
\begin{document}
	
%
\title{On the Capacity of Fractal D2D Social Networks With Hierarchical Communications}

\author{Ying Chen, \IEEEauthorblockN{Rongpeng Li, Zhifeng Zhao, and Honggang Zhang}\\
	\thanks{Y. Chen, R. Li, Z. Zhao, and H. Zhang are with College of Information Science and Electronic Engineering, Zhejiang University. Email: \{21631088chen\_ying, lirongpeng, zhaozf, honggangzhang\}@zju.edu.cn}
	\thanks{This work was supported in part by the Program for Zhejiang Leading Team of Science and Technology Innovation (No. 2013TD20), National Natural Science Foundation of China (No. 61731002, 61701439), Zhejiang Key Research and Development Plan (No. 2018C03056), the National Postdoctoral Program for Innovative Talents of China (No. BX201600133), and the Project funded by China Postdoctoral Science Foundation (No. 2017M610369).}}


%


\maketitle

\begin{abstract}
 The maximum capacity of fractal D2D (device-to-device) social networks with both direct and hierarchical communications is studied in this paper. Specifically, the fractal networks are characterized by the direct social connection and the self-similarity. Firstly, for a fractal D2D social network with direct social communications, it is proved that the maximum capacity is $ \Theta\left(\frac{1}{\sqrt{n\log n}}\right) $  if a user communicates with one of his/her direct contacts randomly, where $ n $ denotes the total number of users in the network, and it can reach up to $ \Theta\left(\frac{1}{\log n}\right) $ if any pair of social contacts with distance $ d $ communicate according to the probability in proportion to $ d^{-\beta} $. Secondly, since users might get in touch with others without direct social connections through the inter-connected multiple users, the fractal D2D social network with these hierarchical communications is studied as well, and the related capacity is further derived. Our results show that this capacity is mainly affected by the correlation exponent $\epsilon$ of the fractal structure. The capacity is reduced in proportional to $ \frac{1}{{\log n}} $ if $ 2<\epsilon<3 $, while the reduction coefficient is $ \frac{1}{n} $ if $ \epsilon=3 $.
\end{abstract}

\begin{IEEEkeywords}
Capacity, D2D Social Networks, Fractal Networks, Hierarchical Social Communications, Self-Similarity
\end{IEEEkeywords}


\section{Introduction}
With the explosive increase of smart devices, social network traffic has witnessed unprecedented growth and imposed huge challenge on traditional content delivery paradigm\cite{Wang2017Propagation}. Emerging as a promising technology to offload the wireless network traffic, device-to-device (D2D) communication allows users in proximity to establish local links and exchange contents directly instead of obtaining data from the cellular base station (BS)\cite{Wu2017Social}. 

Within the D2D communication scenarios, besides the underlying propagation network on the physical layer, the users also form an overlaying social network, where the communication between two users is driven by their social relationship and served by the underlying propagation network. Particularly, with the increasing awareness of security and privacy, trust has become a prerequisite for interactions between mobile users \cite{Guo2017A}. People only communicate with trusted persons rather than geographically close ones. That is to say, the social connection exists if and only if two users trust each other. Depending on whether the two communicating parities have mutual trust or not, social communications in D2D networks can be divided into two major categories:
\begin{itemize}
\item \emph{Direct social communications}: The two end users of communication are mutually trusted and directly linked by a social connection. They are likely to share some kind of intimate relationship, such as families, friends, colleagues and so on.

\item \emph{Hierarchical social communications}: The two end users of communication are not mutually trusted and indirectly connected through a couple of inter-connected users. Rather than the direct communications, they prefer to transfer information via the inter-connected and trusted users.
\end{itemize}

In this work, fractal organization is considered in D2D social networks due to its predominant performance in terms of resilience, scalability and robustness than non-fractal organizations \cite{De2014Mutualistic}. Specifically, a fractal social network can recover quickly from security attacks because the breakdown of a few nodes does not cause the collapse of the whole network. Therefore, it is of significant importance to study fractal D2D social networks and answer the fundamental problem like the capacity of fractal D2D social networks. To be clear, specific practical issues in real world are not the emphases of this work, such as how to design an efficient scheduling algorithm \cite{Wang2017Propagation}, how to stimulate the selfish users to cooperate for social transmissions \cite{Wu2017Social}, or how to establish the trust relationship between two users \cite{Guo2017A}.

The researches on the capacity of a wireless network have continuously aroused intense interests in recent years. However, regardless of its great significance, except the recent works in \cite{Jacquet2013Capacity,Jacquet2015Optimized}, little attention has ever been paid to the capacity of fractal wireless networks, let alone fractal D2D social networks. In this regard, the main focus of this paper is the analysis of the maximum capacity of fractal D2D social networks.

On the other hand, different from some well-known fractal structures with specific geometric shapes as studied in \cite{Jacquet2013Capacity,Jacquet2015Optimized,Maksymyuk2015Fractal}, the fractal social networks in this paper are those with general fractal characteristics. Within the framework of complex networks, the general fractal features have been well studied \cite{Chen1989Fractal,Mandelbrot1983The,Falconer2003Fractal} and are mainly determined by the degree correlations in the connectivity of the networks, which can be characterized by some well-established models or distributions like \cite{Gallos2008Scaling,Song2006Origins,Song2007How}. To be specific, the most vital essence of the general fractal networks studied in this paper is described by the two power-law distributions below:
\begin{itemize}
\item \emph{Joint probability distribution $P(k_1,k_2)$}: It captures the possibility to establish a social connection between two users with degree $ k_{1} $ and $ k_{2} $, and the degree in a fractal D2D social network refers to the number of social connections of a user. As the fundamental requirement for the general fractal social networks \cite{Gallos2008Scaling}, $P(k_1,k_2)$ can be expressed as:
\begin{equation}\label{key1}
P(k_1,k_2)\propto k_1^{-(\gamma-1)}\cdot k_2^{-\epsilon},\ (k_{1}>k_{2})
\end{equation}

\noindent where $ \gamma $ is the degree distribution exponent, $ \epsilon $ is the correlation exponent, and the operator $ \propto $ denotes the proportional relationship between the two sides. This form of expression indicates that $ k_{1} $ and $ k_{2} $ are mutually independent.

\item \emph{Degree distribution $ P(k) $}: Another important characteristic of a general fractal network is known as self-similarity. It has been found that a variety of real complex networks consist of self-repeating patterns on all length scales \cite{Song2005Self}. Self-similarity of a fractal network requires the degree distribution $ P(k) $ to remain invariant when the network grows, namely the so-called scale-free law. In order to meet this requirement, degree distribution must meet Eq. \eqref{key2} below \cite{Albert2002Barab,Newman2003The,Albert1999Internet,Faloutsos1999On}:
\begin{equation}\label{key2}
P(k)\propto k^{-\gamma}.
\end{equation}

 \end{itemize}
 
 Actually, the two power-law distributions above lay the foundations for the capacity analysis of fractal D2D social networks.

\subsection{Related Works}

As a key component of future 5G cellular networks to improve throughput and spectral efficiency, D2D communication has been investigated in many contexts. Particularly, many literatures concerning with the D2D social networks have sprung up. For instance, \cite{Chun2017Device} analyzed the performance of relay-assisted multi-hop D2D communication where the decision to relay was made based on social comparison. In \cite{Feng2017Effective}, the small size social communities were exploited for the resource allocation optimization in social-aware D2D communication. In order to alleviate the security issue in D2D social networks, a secure content sharing protocol was proposed in \cite{Wang2017Secure} to meet the security requirements. 

Among the literatures with regard to D2D social networks, the issue of capacity has barely been considered to our best knowledge. Even so, owing to its fundamental significance, there have existed a great deal of researches on the capacity of various kinds of wireless networks. Philippe Jacquet \emph{et al.} studied the capacity of wireless networks under three models when the emitters and the access point are randomly distributed in an infinite fractal map \cite{Jacquet2013Capacity,Jacquet2015Optimized}. Gupta and Kumar firstly proved that the throughput in ad-hoc wireless networks can reach $ \Theta\left(\frac{1}{\sqrt{n\log n}}\right) $ when the network size is $n$ \cite{Gupta2000The}, where the symbol $ \Theta $ refers to the order of magnitude. The capacity of wireless networks under the relay case was studied in \cite{Gastpar2002Capacity}, and the research on the capacity of hybrid wireless networks was conducted in \cite{Azimdoost2003On}. Kulkarni \emph{et al.} provided a very elementary deterministic approach on the capacity of wireless networks, which gave throughput results in terms of the node locations \cite{Kulkarni2004A}. For social wireless networks, Sadjadpour \emph{et al.} studied the capacity of a scale-free wireless network in which nodes communicate with each other in the context of social groups \cite{Azimdoost2011The}. Particularly, it was discovered that the maximum capacity can be improved in the social scale-free networks compared with the classical conclusion drawn by Gupta and Kumar \cite{Kiskani2013Social,Kiskani2016Effect}. In addition, they studied the capacity of composite networks, namely, the combination of social and wireless ad-hoc networks \cite{Azimdoost2013Capacity}. Bita Azimdoost \emph{et al.} investigated the capacity and latency in an information-centric network when the data cached in each node has a limited lifetime \cite{Azimdoost2016Fundamental}.

In spite of the fact that a vast amount of documents have studied the capacity of various wireless networks, the capacity of fractal networks has been paid little attention to. However, as a vital property of networks, fractal phenomenon has already been discovered in many wireless networking scenarios \cite{Yuan2016Not}. For example, the coverage boundary of the wireless cellular networks shows a fractal shape, and the fractal features can inspire the new design of the hand-off scheme in mobile terminals \cite{Ge2016Wireless,Hao2017Wireless}. Moreover, a large number of significant networks in the real world exhibit the fractal characteristics naturally, such as the world-wide web, yeast interaction, protein homology, and social networks \cite{Strogatz2005Complex,Song2005Self}. In addition, the concept of fractal structure has been taken advantage of in various applications, including the design of antennas for satellite down-link and up-link communications, wireless local area network (WLAN) applications, and other 5G applications \cite{Kaur2016A,Haider2017A}. 

In summary, as one of the most groundbreaking works, this paper studies the capacity of fractal D2D social networks.

\subsection{Contribution}
Distinct from the previous works, the maximum capacity of fractal D2D social networks with both direct and hierarchical interconnections among the users is addressed in this paper. In this regard, some key novel contributions are provided as following:
\begin{itemize}
\item First of all, the capacity of fractal D2D social networks with direct social communications is elaborated. On one hand, it is proved that if a user communicates with one of his/her direct contacts randomly, the maximum capacity is $ \Theta\left(\frac{1}{\sqrt{n\log n}}\right) $. On the other hand, if the two users with distance $ d $ communicate according to the probability in proportion to $ d^{-\beta} $, the maximum capacity can reach up to $ \Theta\left(\frac{1}{\log n}\right) $.

\item Secondly, the relationship between the extendibility of a fractal social network and the correlation exponent $ \epsilon $ is revealed according to the definition in Eq. \eqref{key1} and Eq. \eqref{key2}. It is mathematically proved that $ \epsilon $=3 is the boundary to distinguish whether a fractal network is extensible or not. The fractal network can expand branches continuously when $2<\epsilon\leq3$, while the fractal network stops branching rapidly if $ \epsilon>3 $.

\item Thirdly, the capacity of fractal D2D social networks with hierarchical communications is derived. Compared to the results with direct social communications, it turns out that: 
$ 1) $ If $ 2<\epsilon<3 $, the order of the capacity decreases in proportion to $ \frac{1}{{\log n}} $; $ 2) $ If $ \epsilon=3 $, the capacity reduces in proportion to $ \frac{1}{n} $, which reflects the trade-off between the security and capacity of fractal D2D social networks.

\end{itemize}

The remainder of this paper is organized as follows. The fundamentals of a fractal network as well as the basic knowledge of both direct and hierarchical social communications in fractal D2D networks are introduced, and the corresponding network model is discussed in Section II. Then the maximum throughput with direct social communications is derived in Section III. Afterwards, the deductions are extended to the case with hierarchical social communications in Section IV. Numerical simulation results are discussed in Section V. Finally, a conclusion is drawn in Section VI.

\section{Background and Models}
\subsection{Fundamentals of A Fractal Network}
In order to characterize the general fractal networks rather than the geometric fractal ones studied in \cite{Jacquet2013Capacity,Jacquet2015Optimized}, it is essential to introduce the basic concept of renormalization through the box-covering algorithm \cite{Song2005Self,Song2006Origins,Song2007How}. 

As illustrated in Fig. 1, renormalization is a technique to examine the internal relationship among the nodes in a complex network by using a box to cover several nodes and virtually replacing the whole box by a new representative node. Besides, if there exists a link between any two nodes in two boxes respectively, then the two corresponding representative nodes evolved from the boxes will be connected. 
\begin{figure}[htbp]
	\centering
	\includegraphics[scale=0.4]{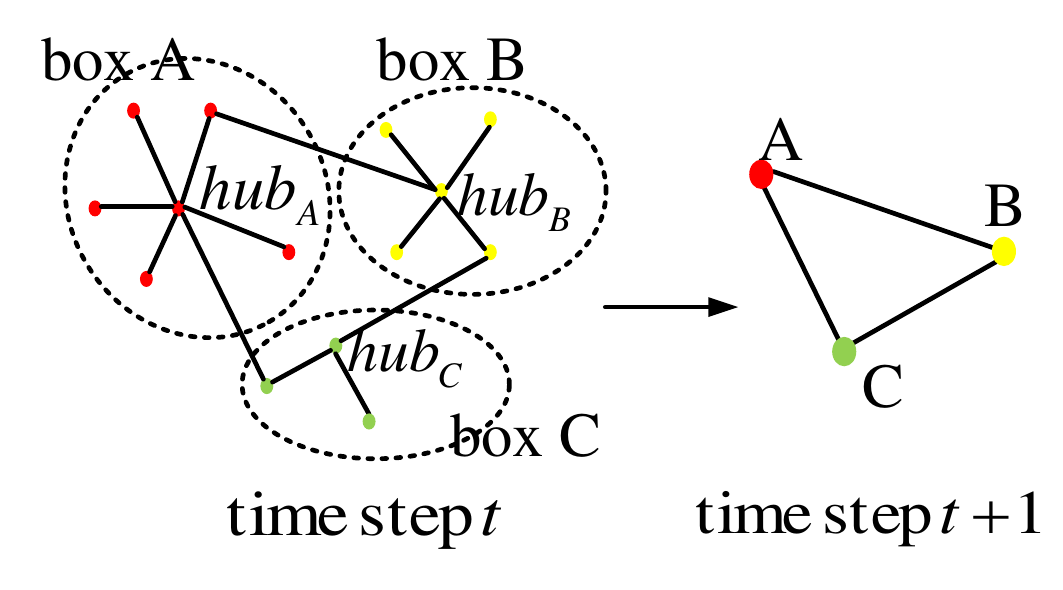}
	\caption{The illustration of renormalization of a fractal network.}
\end{figure}

Mathematically, the network can be minimally covered by $ N_{B}(l_{B}) $ boxes of the same length scale $ l_{B} $ under renormalization, where $ l_{B} $ is the size of the box measured by the maximum path length between any pair of nodes inside the box, and $ N_{B}(l_{B}) $ is the minimum value among all possible situations. To be specific, the size of boxes $ l_{B} $ in Fig. 1 is 2 and the number of boxes $ N_{B}(l_{B}) $ is 3 (box A, B, and C). 

In essential, if the network is a general fractal network, the following relations hold, namely \cite{Song2006Origins,Cohen2003Scale}:
\begin{equation}\label{fractal}
\left\{ {\begin{array}{*{20}{c}}
	\begin{aligned}
	&{{N_B}({l_B})/n\propto{l_B}^{ - {d_B}}}\\
	&{{k_B}({l_B})/{k_{hub}}\propto {l_B}^{ - {d_g}}}\\
	&{{n_h}({l_B})/{k_B}({l_B})\propto {l_B}^{ - {d_e}}},
	\end{aligned}
	\end{array}} \right.
\end{equation}
	
\noindent where $ n $ is the number of nodes in the network. A hub indicates the node with the largest degree inside each box, while $ k_{B}(l_{B}) $ and $ k_{hub} $ denote the degree of the box and the hub respectively. $ n_{h}(l_{B}) $ refers to the number of links between the hub of a box and the nodes in other connected boxes. Take box A in Fig. 1 for example, the variables $ k_{B}(l_{B}) $, $ k_{hub} $ and $ n_{h}(l_{B}) $ are 2, 6 and 1, respectively. The three indexes $ d_{g} $, $ d_{B} $ and $ d_{e} $ indicate the degree exponent, the fractal exponent, and the anti-correlation exponent, respectively.   

Viewing the above process of renormalization from the perspective of time steps, the number of boxes $ {N_B}({l_B}) $ in this time step is also the number of nodes in the next time step, and the first equation in Eq. \eqref{fractal} indicates that the ratio of the numbers of nodes between successive time steps is proportional to $ {l_B}^{ - {d_B}} $, and the power-law relationship remains unchanged under renormalization, obeying the scale-free law. In the similar way, the degree of the box $ k_{B}(l_{B}) $ at present is the degree of the hub $ k_{hub} $ in the next time step as well, and the second equation in Eq. \eqref{fractal} shows that the ratio of the hub degrees between successive time steps is a scale-invariantly exponential function of the box size  $ l_{B} $. Therefore, what the first and second equations in Eq. \eqref{fractal} reveal is actually the topological self-similarity of fractal networks \cite{Song2006Origins}. In addition, the degree exponent $ d_{g} $ and the fractal exponent $ d_{B} $ are both finite for fractal networks.

Moreover, the ratio $ {n_h}({l_B})/{k_B}({l_B}) $ in the third equation in Eq. \eqref{fractal} reveals the contribution of hub nodes in box-box connections, and it decreases sharply with the increase of the length scale $ l_{B} $. Actually, this equation illustrates the hub repulsion phenomenon, i.e., a node with a large degree prefers not to be linked to another node with a large degree, which is another essential property of fractal networks. Therefore, the anti-correlation exponent $ d_{e} $ reveals the repulsion effect between the hubs, and large $ d_{e} $ tends to result in a fractal networking structure. As shown in Fig. 1, instead of establishing a connection with another hub, each hub node prefers to connect with a non-hub node in another box.

In \cite{Gallos2008Scaling}, it has been proved that there exist certain relations among the aforementioned key parameters: $ \gamma  = 1 + \frac{{{d_B}}}{{{d_g}}} $ and $ \epsilon  = 2 + \frac{{{d_e}}}{{{d_g}}} $, which suggest that $ \gamma $ and $\epsilon $ in a fractal wireless network are larger than $ 1 $ and $ 2 $, respectively. Also the degree distribution exponent is usually in the range $ 2<\gamma<3 $ in real complex networks \cite{Song2005Self}. Please note that for concision, hereinafter all the following relevant mathematic deductions will be characterized by the key parameters $ \gamma $ and $ \epsilon $, instead of $ d_{B} $, $ d_{g} $ and $ d_{e} $.

\subsection{Knowledge of Both Direct and Hierarchical Social Communications in Fractal D2D Social Networks}
Fig. 2(a) illustrates the direct/$ level $-1 social communications in a fractal D2D social network. As we can see, four users, namely Bob, Jane, Joy and Rose, are directly connected with Alice and are regarded as the direct, or $ level$-1 contacts of Alice. If Alice chooses to communicate with Bob among her four direct contacts, then Alice and Bob are known as the source user and the destination user, respectively. Usually, a user has more than one direct contacts, and the degree $ k $ refers to the number of his/her $ level$-1 contacts. In the case of $ level$-1 social communications, the degree distribution and the joint probability distribution are the aforementioned $P(k)$ and $P(k_1,k_2)$, respectively. As discussed later in Section II-C, the direct/$ level $-1 contact does not imply there physically exist some direct links. Instead, the pair of users for direct social contact might have to rely on some relaying nodes in the underlying physical propagation network.

In addition to the direct case, the social communications in fractal D2D social networks can actually be hierarchical as depicted in Fig. 2(b). If Alice wants to get in touch with Victoria who she does not trust, the data packets have to be transmitted through the inter-users Bob and Jack. That is to say, a source user can communicate with one of his/her $ level$-$L\ (L=1,2,\cdot \cdot \cdot,L_{max}) $ contacts through $ L-1 $ inter-users to make sure that every transmission is carried out between two users with mutual trust, and $ L_{max} $ refers to the maximum social relationship level. For instance, in the case of $ level$-2 social communications, Jack is indirectly connected with the source user Alice through one inter-user Bob, so Jack is one of the $ level $-2 contacts of Alice, and he can be selected as the destination user among all the $ level $-2 contacts to communicate with Alice. Similarly, Victoria is referred to as one of the $ level $-3 contacts of Alice, and so on.

\begin{figure}[htbp]
	\centering
	\includegraphics[scale=0.3]{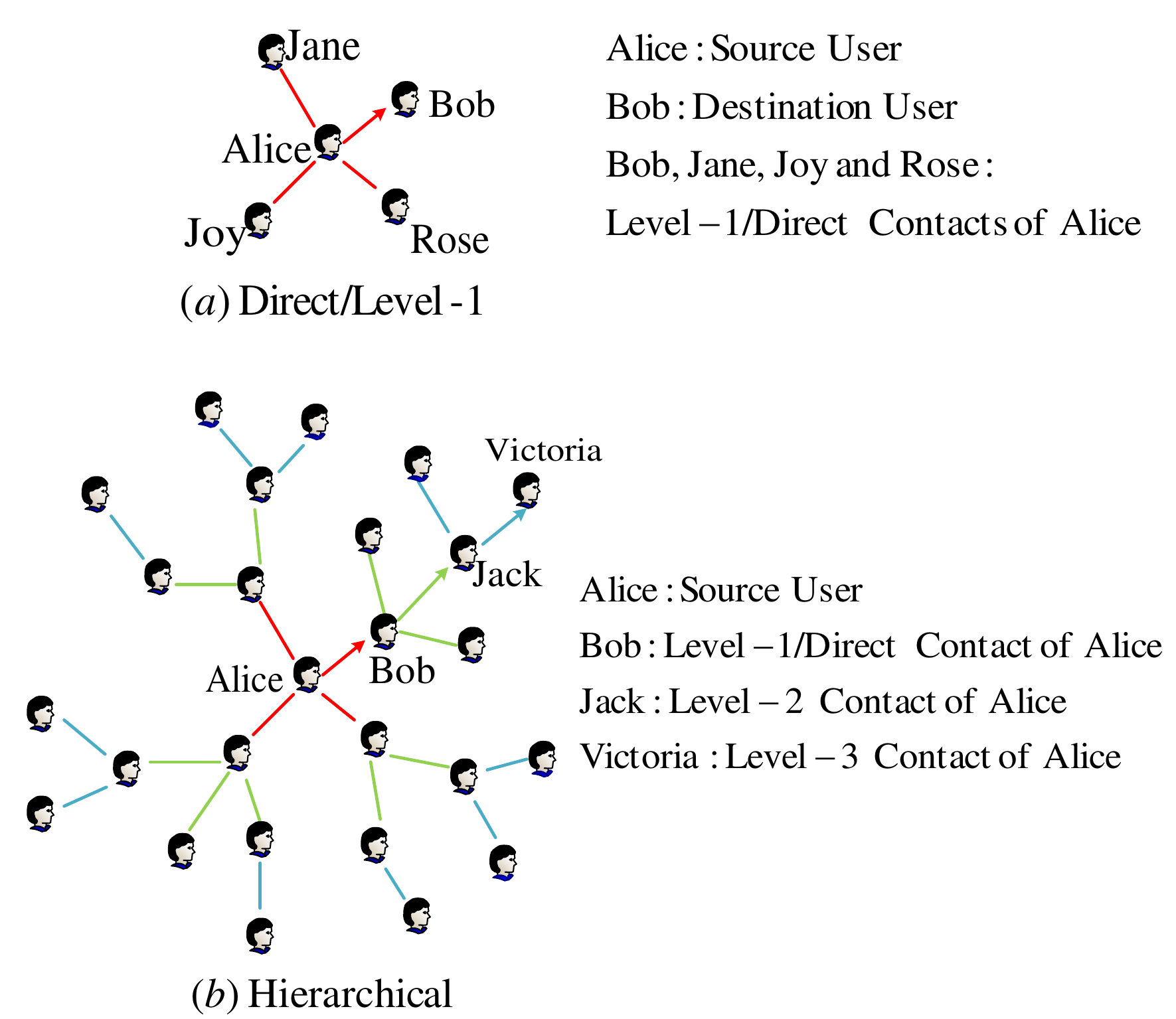}
	\caption{(a) Direct/$ Level $-1 social communications in a fractal D2D social network; (b) Hierarchical social communications in a fractal D2D social network.}
\end{figure}

To enhance the understanding of mathematical derivations in Section IV, it is necessary to introduce the concept of $ level $-$ L $ graphs of the social topology. Without loss of generality, the superscript $ L $ is used to denote a $ level $-$ L $ case. If all $ level $-2 contact pairs in Fig. 3(a) are connected virtually, then a new graph is obtained as shown in Fig. 3(b). In $ level $-2 graph, the degree distribution is defined as $P(k^{(2)})$, where the degree $ k^{(2)} $ of a user is the $ level $-2 degree, and refers to the number of his/her $ level $-2 contacts or his/her links in the $ level $-2 graph. In the same way, the $ level $-3 graph can be obtained in Fig. 3(c) by connecting all $ level $-3 contact pairs virtually, and the degree distribution here is defined as $P(k^{(3)})$, where the degree $ k^{(3)} $ of a user is the $ level $-3 degree, and so forth. 
\begin{figure}[htbp]
	\centering
	\includegraphics[scale=0.3]{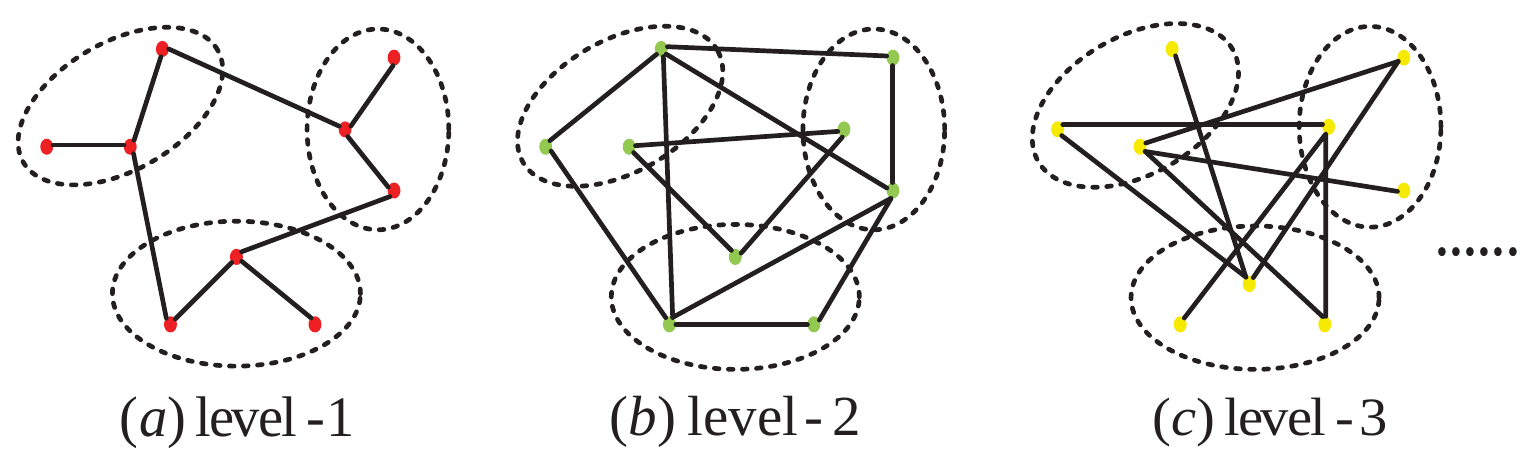}
	\caption{The $ level $-$ L $ graphs of the social topology by connecting all $ level $-$ L $ contact pairs virtually.}
\end{figure}

\subsection{Network Model}
In order to clarify the capacity of the above fractal D2D social networks with both direct and hierarchical communications clearly and orderly, it is assumed that all the $ n $ users are uniformly distributed in a unit area square. Also the fractal D2D social network is treated as a static network because the users barely move during one transmission frame.

\begin{figure}[htbp]
	\centering
	\includegraphics[scale=0.18]{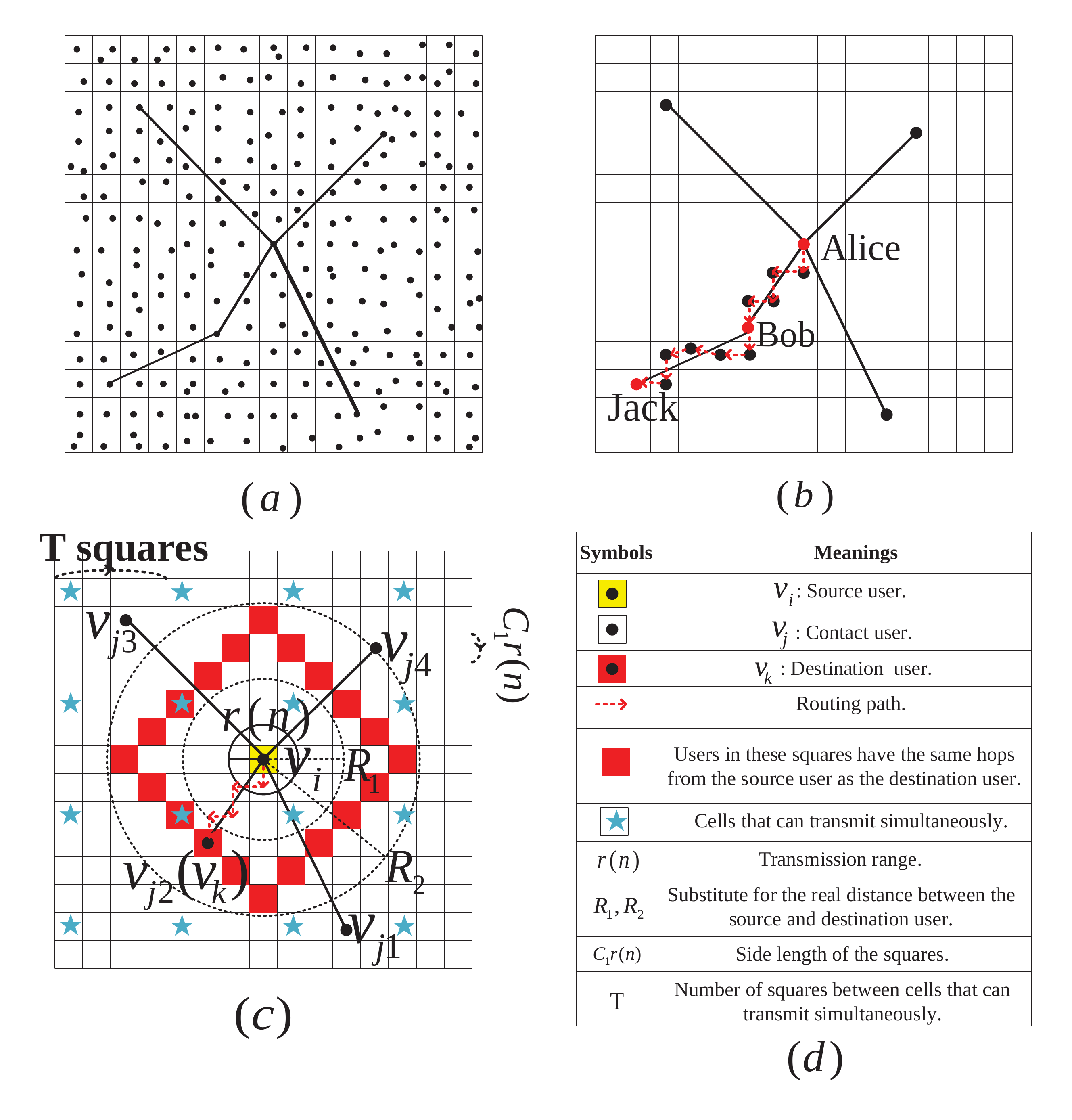}
	\caption{(a) An illustrative part of the overlaying fractal D2D network with social interconnections; (b) The underlying physical propagation network serves to forward data for a transmission via multi-hop routing between any pair of social contacts; (c) The fractal D2D social network deployed in a standard unit area square model \cite{Kulkarni2004A,Kiskani2016Effect}; (d) The table of the symbols in the model.}
\end{figure}

All the potential users form an underlying D2D propagation network on the physical layer, as well as an overlaying fractal social network from the viewpoint of social connections. An illustrative part of the overlaying fractal D2D social network is shown in Fig. 4(a), and the connection between two users stands for the relationship of mutual trust. It is noteworthy that the topological fractal social network is formed by the D2D social connections of all the involved users following the aforementioned degree distributions $ P(k) $ and $ P(k_{1},k_{2}) $, which is not contradictory with the general assumption of physically uniformly distributed users.

As depicted in Fig. 4(b), the underlying D2D physical propagation network has to be distinguished from the overlaying fractal social network, where the propagation network serves the social communications and forwards data for a transmission via multi-hop routing between any pair of social contacts. For example, when Alice wants to communicate with Jack, she has to get in touch with Jack through Bob, as we discussed in Section II-B. However, Alice and Bob cannot exchange data directly even though they are socially connected because they are not physically close enough to exchange contents locally. In order to transmit a packet from Alice to Bob, a few other nodes in the underlying D2D propagation network have to serve as relay nodes as the red dotted path in Fig. 4(b) shows, so does the transmission from Bob to Jack. It has been explained in \cite{Kiskani2016Effect} that the relay nodes will never cause traffic bottleneck, so the underlaying propagation network will not change the capacity of the overlaying social network. Please note that this kind of multi-hop data relaying is complying with the same approach as widely used in \cite{Kulkarni2004A,Azimdoost2011The,Kiskani2013Social,Kiskani2016Effect}.

The representative case of $ level $-1 social communications is described in Fig. 4(c), and the symbols in the model are listed in Fig. 4(d). The destination user $ v_{k} $ is chosen among the four $ level $-1 contacts (user $ v_{j} $). In $ level$-$L\ (L=2,3,...,L_{max}) $ situations, the pattern is almost the same except that the destination user is re-selected among the $ level $-$ L $ contacts. 

Similar to the approaches widely used in \cite{Kulkarni2004A,Azimdoost2011The,Kiskani2013Social,Kiskani2016Effect}, a simple multi-hop routing scheme in the physical space domain is adopted here. When the source user is about to send a data packet, it chooses one user closest to its destination user from its neighboring squares to relay the packet. This kind of physical relaying steps keep going until the data packet eventually reaches the destination user after multiple hops. The red dotted line with arrows denotes one possible routing path in Fig. 4(c). The data packet can be successfully transported for any pair of transmission under the condition that there is at least one user in each square. Surely this condition can be satisfied with a probability approaching $ 1 $ according to the classical work in \cite{Kulkarni2004A}. In Fig. 4(c), since every hop transmits the packet from one small square to one of its neighboring squares, all the squares marked in red solid have the same hops $ x $ from the source user as the destination user $ v_{k} $, and the total number of hops is $ 4x $. The radius $ R_{1} $ and $ R_{2} $ of the two dotted-line circles are used as the indicative distances between the source and destination user instead of their real distance.

Corresponding to the D2D social communication scenario, the widely employed protocol model in \cite{Kulkarni2004A,Kiskani2016Effect,Feng2006Scaling} is adopted as the measurement of a successful physical transmission. Firstly, as mentioned above, two users can exchange contents directly only when they are geographically close enough in the D2D communication, so an upper bound of the distance has to be set between two users who can physically communicate directly. Secondly, interference is one of the main issues to be paid attention to in the D2D communication. In order to rule out the influence of interference, the distance between the transmitter and the receiver has to meet some lower bound. According to the above protocol model, a physical transmission is successful if and only if the Euclidean distance between two users meets the conditions: $ |X_{i}-X_{j}|\leqslant r(n) $ and $ |X_{k}-X_{i}|\geq (1+\Delta)|X_{i}-X_{j}| $, where $ X_{i} $ and $ X_{j} $ refer to the transmitter and the receiver respectively, $ X_{k} $ denotes any other transmitter sharing the same channel with $ X_{i} $ and $ \Delta $ is the guard zone factor. It has been proved that the transmission range $ r(n) $ must reach $ \Theta\left(\sqrt{\frac{\log n}{n}}\right) $ to guarantee the connectivity of the network \cite{Penrose1997The}. In Fig. 4(c), the  solid-line circle with the radius of $ r(n) $ displays the transmission range.

For keeping consistence with the above protocol model in analyzing the capacity, similarly to \cite{Gupta2000The,Kulkarni2004A}, a TDMA (Time Division Multiple Access) scheme is designated as the MAI (multiple access interference) avoidance method. As shown in Fig. 4(c), the networking area is divided into a number of smaller squares with side length $ C_{1}r(n) $, where $ C_{1} $ is a constant. Equivalent to the condition $ |X_{k}-X_{i}|\geq (1+\Delta)|X_{i}-X_{j}| $ in the classical protocol model, the interference units refer to those squares containing at least two nodes closer than $ (2+\Delta)r(n) $ respectively \cite{Gupta2000The}, and these squares which can simultaneously transmit data packets should not be the interference units with each other. Therefore, users in the squares signed with blue stars in Fig. 4(c), which are at least $ T $ squares away from each other, are permitted to transmit data packets at the same time, where $ T\geq (2+\Delta)/C_{1} $. 

Since the selection rule of the destination user affects the capacity of the fractal D2D social network as well, the destination user is chosen in two different ways in this paper. In the first case, the destination user is selected according to the uniform distribution. That is to say, a user communicates with one of his/her contacts randomly, and all the potential contacts have the same opportunity to communicate with the user. In the second case, the destination user is selected according to the power-law distribution $ d^{-\beta} $ \cite{Latane1995Distance}, where $ d $ refers to the distance between the source and destination user, and $ \beta $ is the frequency parameter. This selection rule is considered to be reasonable as Latane \emph{et al.} \cite{Latane1995Distance} discovered that a user prefers to communicate with physically closer user among his/her social contacts, and the probability is proportional to the power-law of the distance. In other words, the social contacts closer to the source user have more opportunities to communicate with him/her. 
    
For simplicity of representation, some essential definitions are given and the relationship between them is highlighted as following.

$ \mathbf{Definition\ 1.} $ The elementary symmetric polynomial \cite{JIPAM2003New} $ \sigma_{p,N}(Q')$, $1\leq q\leq N $  of variables $ Q'=(q_{1},q_{2},...,q_{N}) $ is noted as
\[ 
\begin{aligned}
&\sigma_{p,N}(Q')=\sigma_{p,N}(q_{1},q_{2},...,q_{N})\\
&\qquad\qquad=\sum\nolimits_{ 1\leq i_{1}\leq i_{2}\leq...\leq i_{p}\leq N}q_{i_{1}}q_{i_{2}}... q_{i_{p}}.
\end{aligned}
 \]

$ \mathbf{Definition\ 2.} $ The elementary symmetric polynomial \cite{JIPAM2003New} $  \sigma^{\overline{k}}_{p,N-1}(Q') $, $ 1\leq p\leq N-1 $ of variables $ Q'=(q_{1},q_{2},\ldots,q_{N}) $ except $ q_{k} $ is noted as
\[
\sigma^{\overline{k}}_{p,N-1}(Q')=\sigma_{p,N-1}(q_{1},q_{2},...,q_{k-1},q_{k+1},...,q_{N}).
\]

From \cite{Kiskani2016Effect,JIPAM2003New}, we can have the following lemma.

 $ \mathbf{Lemma\ 1.} $ Let the set $ Q' = \{ {q_1},{q_2},...,{q_N}\}  $ contains $ N\geq 2 $ non-negative real numbers. If $ q $ is finite, then we have \cite{JIPAM2003New}: 
\[ 
	\frac{\sigma_{1,N}(Q')\cdot\sigma_{q,N}(Q')}{(q+1)\cdot\sigma_{q+1,N}(Q')}=\Theta(\frac{N}{N-q}).
\]

To be clear, it turns out that the symbol $ \Theta $ is not about the numerical value, it is about the speed of growth. In other words, two variables on the two sides of an equation have the same speed of growth.

\section{The Upper Bound of the Capacity With Direct Social Communications}

In this section, the aforementioned properties of fractal D2D social networks are followed and the specific derivation procedure of the maximum capacity with direct social communications is clarified. That is to say, only the $ level $-1 social communication is considered in this section, and all the contacts, degrees, degree distribution or joint probability distribution in this section are default $ level $-1. In addition, the impact of a particular destination selection rule on the maximum achievable throughput is studied by taking account of two different cases, including uniformly and power-law distributed destinations.  

For the convenience for understanding, a list of all the symbols and their explanations is shown in Table I.

\begin{table}[htbp]
	\centering  
	\caption{The list of the symbols and their meanings in Section III.}
	\begin{tabular}{|c|c|}
		\hline
		Symbols &Meanings \\ \hline  
		$ n $ &The total number of users.\\  
		\rowcolor{mygray}      
		$ k $ &The degree of a user.\\      
		$ q $ &The degree of the source user.\\ 
		\rowcolor{mygray} 
		$ q_{0} $ &A relative large degree to divide the\\
		\rowcolor{mygray} 
		&range of degrees into two parts.\\ 
		$ N $ &The number of potential contacts \\
		&whose degree is less than $ q $.\\ 
		\rowcolor{mygray} 
		$ \gamma $ &The degree distribution exponent \\
		\rowcolor{mygray} 
		&of a fractal network.\\ 
		$ \epsilon $ &The correlation exponent of \\
		&a fractal network.\\ 
		\rowcolor{mygray} 
		$ M_{\gamma,\epsilon} $ &The normalization constant in the \\
		\rowcolor{mygray} 
		&joint probability distribution $ P(k_{1},k_{2}) $.\\ 
		$ \mathbf{C} $ &The set of all $ level $-1 contacts.\\ 
		\rowcolor{mygray} 
		$ v_{i} $ &The source user.\\ 
		$ v_{t} $ &The destination user.\\ 
		\rowcolor{mygray} 
		$ v_{k} $ &A particular contact who is \\
		\rowcolor{mygray} 
		&selected as the destination user.\\ 
		$ q_{k} $ &The degree of $ v_{k} $.\\ 
		\rowcolor{mygray} 
		$ \lambda $ &The data rate for every user.\\ 
		$ \lambda_{max} $ &The maximum achievable capacity.\\ 
		\rowcolor{mygray} 
		$ X $ &The number of hops from the \\
		\rowcolor{mygray} 
		&source user to the destination user.\\ 
		$ E[X] $ &The average number of hops.\\ 
		\rowcolor{mygray} 
		$ E_{1} $ &The average number of hops when \\
		\rowcolor{mygray} 
		&$ q \le {q_0} $ under the case of uniformly \\
		\rowcolor{mygray} 
		&distributed destinations.\\ 
		$ E_{2} $ &The average number of hops when \\
		&$ q > {q_0} $ under the case of uniformly \\
		&distributed destinations.\\ 
		\rowcolor{mygray} 
		$ s_{l} $ &The red squares in Fig. 4(c), \\
		\rowcolor{mygray} 
		&where $ l=1,2,\ldots 4x $.\\ 
		$ d $ &The distance between the source user \\&and the destination user.\\ 
		\rowcolor{mygray} 
		$ d_{j} $ &The distance between the source user \\
		\rowcolor{mygray} 
		&and his $ j $-th contact.\\ 
		$ \beta $ &The frequency parameter.\\ 
		\rowcolor{mygray} 
		$ E_{3} $ &The average number of hops when \\
		\rowcolor{mygray} 
		&$ q \le {q_0} $ under the case of power-law \\
		\rowcolor{mygray} 
		&distributed destinations.\\ 
		$ E_{4} $ &The average number of hops when \\
		&$ q > {q_0} $ under the case of power-law \\
		&distributed destinations.\\ \hline
	\end{tabular}
\end{table}

\subsection{The Case of Uniformly Distributed Destinations}
In the first case, the uniform distribution of the destination users is considered. In other words, the source user selects one of his/her $ level $-1 contacts as the destination user randomly. In this situation, the result of the maximum capacity is given in Theorem 1 and proved afterwards. It is noteworthy that the proof may seem similar as \cite{Kiskani2016Effect}, but actually totally different in detail because the capacity of fractal networks is focused on here.

$ \mathbf{Theorem\ 1.} $ For a fractal D2D social network with $ n $ users satisfying the conditions below: 1) the $level $-1 social contacts are selected according to the joint probability distribution $ P({k_1},{k_2}) = \frac{{k_1^{ - (\gamma  - 1)}k_2^{ - \epsilon }}}{{{M_{\gamma ,\epsilon }}}},\ (k_1>k_2) $, where $ M_{\gamma ,\epsilon } $ is the normalization constant; 2) the $level $-1 degree of each user follows the power-law degree distribution $ P(k) = \frac{{{k^{ - \gamma }}}}{{\sum_{k = 1}^n {{k^{ - \gamma }}} }} $; 3) the destination user $ v_{t} $ is chosen by the source user $ v_{i} $ according to the uniform distribution $ P({v_t} = {v_k}|{v_k} \in {\bf{C}}) = \frac{1}{q} $, where $ q $ is the $level $-1 degree of the source user, $ \bf{C} $ is the set of all $level $-1 contacts, and $ v_{k} $ is a particular contact who is selected as the destination user. Then the maximum capacity $ \lambda_{max} $ of the fractal D2D social network with direct social communications is
\begin{equation}\label{Theorem1}
{\lambda _{\max }} = \Theta\left(\frac{1}{{\sqrt {n \cdot \log n} }}\right),
\end{equation}

\noindent where the symbol $ \Theta $ refers to the order of magnitude.

Before giving the proof, some key lemmas are listed as follows. First, from \cite{Kiskani2016Effect}, we can have the following lemma.

$ \mathbf{Lemma\ 2.} $ Assume that $ \lambda $ is the data rate for every user, $ \lambda_{max} $ is the maximum capacity of the fractal D2D social network. $ X $ is the number of hops from the the source user to the destination user. $ E[X] $ denotes the expectation of $ X $ for any social transmission pair. Then we have
\begin{equation}\label{Lemmma2}
\lambda\leqslant\lambda_{max}=\Theta\left(\frac{1}{\log n\cdot E[X]}\right).
\end{equation}

$ \mathbf{Lemma\ 3.} $ Let the degree of the source user be $ q $, where $ q=1,2,...,n $. $ v_{k} $ is a particular contact who is selected as the destination user, and $ q_{k}$ is the degree of $ v_{k} $. The variables $ q_{i_{1}},q_{i_{2}}, \cdot \cdot \cdot ,q_{i_{N}} $ in $ Q = (q_{i_{1}}^{ - \epsilon },q_{i_{2}}^{ - \epsilon }, \cdot  \cdot \cdot,q_{i_{N}}^{ - \epsilon }) $ denote the degrees of $ N $ potential $ level $-1 social contacts whose degree is smaller than $ q $. $ x $ is the number of hops from the source to the destination user, and $ s_{l}\ (l=1,2,\ldots 4x) $ stands for the red square in Fig. 4(c). Then the average number of hops is 
\begin{equation}\label{Lemma3}
E[X] = \sum\limits_{q = 1}^n {\frac{{{q^{ - \gamma }}}}{{\sum_{b = 1}^n {{b^{ - \gamma }}} }}}  \cdot \sum\limits_{x=1}^{\frac{1}{{r(n)}}} {x\sum\limits_{l = 1}^{4x} {\sum\limits_{{v_k} \in {s_l}} {\frac{{q_k^{ - \epsilon }\cdot\sigma _{q - 1,N - 1}^{\overline k }(Q)}}{{{q\cdot \sigma _{q,N}}(Q)}}} } }.
\end{equation}

The proof is left in Appendix A.

Next, $ E[X] $ is divided into two separate cases $ E_{1} $ and $ E_{2} $ with a boundary $ q_{0} $, which is a constant and indicates a relatively large degree. $ E_{1} $ is the average number of hops when $ q \le {q_0} $, where the degree $ q $ of the source user is a finite integer, meanwhile $ E_{2} $ is the average number of hops when $ q > {q_0} $, where $ q $ is considered to be infinite.

$ \mathbf{Lemma\ 4.} $ When the degree of the source user is not greater than $ q_{0} $, i.e., $ q \le {q_0} $, the average number of hops $ E_{1} $ is
\begin{equation}\label{Lemma4}
E_{1} =\Theta \left(r{{(n)}^{ - 1}}\right). 
\end{equation}

\emph{Proof}: According to the meaning of $ E_{1} $, it can be given as
\begin{equation}\label{Lemma4.1}
{E_1} = \sum\limits_{q = 1}^{{q_0}} {\frac{{{q^{ - \gamma }}}}{{\sum_{b = 1}^n {{b^{ - \gamma }}} }}}  \cdot \sum\limits_{x=1}^{\frac{1}{{r(n)}}} {x\sum\limits_{l = 1}^{4x} {\sum\limits_{{v_k} \in {s_l}} {\frac{{q_k^{ - \epsilon }\cdot\sigma _{q - 1,N - 1}^{\overline k }(Q)}}{q\cdot{{\sigma _{q,N}}(Q)}}} } } .
\end{equation}

All situations of selecting $ q-1 $ users from $ \bf{C} $ can be parted into two categories according to the condition whether $ {v_k} $ is chosen or not. If it is chosen, other $ q-2 $ users have to be chosen from $ \bf{C} $ besides $ {v_k} $. Otherwise $ q-1 $ users are chosen in $ \bf{C} $ except $ {v_k} $. That is to say,
\[\begin{aligned}
&\sigma _{q - 1,N - 1}^{\overline k }(Q) = {\sigma _{q - 1,N}}(Q) - q_k^{ - \epsilon }\cdot\sigma _{q - 2,N - 1}^{\overline k }(Q){\rm{ }}\\
&= {\sigma _{q - 1,N}}(Q) - q_k^{ - \epsilon }\left({\sigma _{q - 2,N}}(Q) - q_k^{ - \epsilon }\cdot\sigma _{q - 3,N - 1}^{\overline k }(Q)\right).
\end{aligned}\]

Since every term above is positive, then we have
\begin{equation}\label{Lemma4.2}
{\sigma _{q - 1,N}}(Q) - q_k^{ - \epsilon } \cdot {\sigma _{q - 2,N}}(Q) < \sigma _{q - 1,N - 1}^{\overline k }(Q) < {\sigma _{q - 1,N}}(Q).
\end{equation}

According to Eq. \eqref{Lemma4.2}, it turns out the upper bound and the lower bound have the same order:
\begin{equation}\label{Lemma4.3}
E_{1} = {\rm O} \left(r{{(n)}^{ - 1}}\right),\ E_{1} = {\rm \Omega} \left(r{{(n)}^{ - 1}}\right).
\end{equation}

The proof of Eq. \eqref{Lemma4.3} in detail is in Appendix B and now Lemma 4 is proved. $\hfill\blacksquare$

$ \mathbf{Lemma\ 5.} $ When the degree of the source user is greater than $ q_{0} $, i.e., $ q>{q_0} $, the average number of hops $ E_{2} $ is
\begin{equation}\label{Lemma5}
E_{2} =\Theta \left(r{{(n)}^{ - 1}}\right). 
\end{equation}

The proof is left in Appendix C.

Now Theorem 1 can be proved. 

\emph{Proof}: In order to get the result in Theorem 1, the proof sketch below is followed inspired by \cite{Kiskani2016Effect}. Firstly, the relationship between the capacity and the average number of hops $ E[X] $ is presented in Lemma 2, so the problem can be solved by finding out $ E[X] $. Secondly, the expression of $ E[X] $ is given in Lemma 3. Thirdly, $ E[X] $ is separated into two cases $ E_{1} $ and $ E_{2} $ according to the boundary $ q_{0} $, and $ E_{1} $ and $ E_{2} $ are obtained respectively. Therefore, the capacity derivation can be achieved by backtracking.  Based on the results in Lemma 4 and Lemma 5, 
\begin{equation}\label{9}
E[X] = {E_1} + {E_2} = \Theta \left(r{{(n)}^{ - 1}}\right).
\end{equation}

Together with Lemma 2, the results in Theorem 1 is obtained. $\hfill\blacksquare$ 

\subsection{The Case of Power-law Distributed Destinations}
In the second case, it is assumed that the probability that the source user communicates with one of his/her $ level $-1 contacts is proportional to $ d^{-\beta} $, where $ d $ refers to the distance between two users and $ \beta $ indicates that the closer contacts have more opportunities to communicate with the source user. The fractal D2D social network achieves another maximum throughput in this situation, which is clarified in Theorem 2. 

$ \mathbf{Theorem\ 2.} $ For a fractal D2D social network with $ n $ users satisfying the conditions below: 1) the social contacts are selected according to the joint probability distribution $ P({k_1},{k_2}) = \frac{{k_1^{ - (\gamma  - 1)}k_2^{ - \epsilon }}}{{{M_{\gamma ,\epsilon }}}},\ (k_1>k_2) $, where $ M_{\gamma ,\epsilon } $ is the normalization constant; 2) the degree of each user follows the power-law degree distribution $ P(k) = \frac{{{k^{ - \gamma }}}}{{\sum_{k = 1}^n {{k^{ - \gamma }}} }} $; 3) the destination user $ v_{t} $ is chosen according to the power-law distribution $ P({v_t} = {v_k}|{v_k} \in {\bf{C}}) = \frac{{{d^{ - \beta }}}}{{\sum_{j = 1}^q {d_j^{ - \beta }} }} $, where $ \bf{C} $ is the set of all contacts, $ v_{k} $ is a particular contact who is selected as the destination user, $ d $ is the distance from the source user to the destination user, $ d_{j} $ is the distance between the source user and his $ j $-th contact, and $ \beta $ is the frequency parameter. Then the maximum capacity $ \lambda_{max} $ of the fractal D2D social network is
\begin{equation}\label{Theorem2}
{\lambda _{\max }} = \left\{ {\begin{array}{*{20}{c}}
	\begin{aligned}
	&{\Theta\left(\frac{1}{{\sqrt {n \cdot \log n} }}\right),{\rm{     \ \quad \qquad\qquad 0}} \le \beta  \le {\rm{2}}};\\
	&{\Theta\left(\frac{1}{{\sqrt {{n^{3 - \beta }} \cdot \log {n^{\beta  - 1}}} }}\right),{\rm{    \ \qquad 2}} < \beta < {\rm{3}}};\\
	&{\Theta\left (\frac{1}{{\log n}}\right),{\rm{        \qquad\qquad\qquad\qquad }}\beta \geq 3}.
	\end{aligned}
	\end{array}} \right.
\end{equation}

The proof of Theorem 2 is pretty similar to Theorem 1, so only the relevant key lemmas in the derivation procedure are given. 

$ \mathbf{Lemma\ 6.} $ Let the degree of the source user be $ q $, where $ q=1,2,\ldots,n $. $ v_{k} $ is a particular contact who is selected as the destination user. $ q_{k}$ is the degree of $ v_{k} $ and $ d_{k} $ is the distance from the source user to $ v_{k} $. The variables $ q_{i_{1}},q_{i_{2}}, \cdot \cdot \cdot ,q_{i_{N}} $ in $ Q = (q_{i_{1}}^{ - \epsilon },q_{i_{2}}^{ - \epsilon }, \cdot  \cdot \cdot,q_{i_{N}}^{ - \epsilon }) $ denote the degrees of $ N $ potential $ level $-1 social contacts whose degree is smaller than $ q $. Let $ D = (d_1^{ - \beta },d_2^{ - \beta }, \cdot  \cdot  \cdot d_q^{ - \beta }) $, where $ {d_j}\ (1\le j\le q) $ in $ D $ denotes the distance between the $ j $-th social contact and the source user. Then the average number of hops is 
\begin{equation}\label{Lemma6}
\begin{aligned}
&E[X]=\sum\limits_{q = 1}^n {\frac{{{q^{ - \gamma }}}}{{\sum_{b = 1}^n {{b^{ - \gamma }}} }}}  \cdot\\
&\qquad\sum\limits_{x=1}^{\frac{1}{{r(n)}}} {x\sum\limits_{l = 1}^{4x} {\sum\limits_{{v_k} \in {s_l}} {\frac{{q_k^{ - \epsilon }\cdot\sigma _{q - 1,N - 1}^{\overline k }(Q)}}{{{\sigma _{q,N}}(Q)}}} } }  \cdot \frac{{d_k^{ - \beta }}}{{{\sigma _{1,q}}(D)}}.
\end{aligned}
\end{equation}

$ \mathbf{Lemma\ 7.} $ When the degree of the source user is not greater than $ q_{0} $, i.e., $ q \le {q_0} $, the average number of hops $ E_{3} $ is
\begin{equation}\label{Lemma7}
E_{3} = \left\{ {\begin{array}{*{20}{c}}
	\begin{aligned}
	&{\Theta \left(r{{(n)}^{ - 1}}\right),{\rm{      \ \ \qquad 0}} \le \beta  \le {\rm{2}}};\\
	&{\Theta\left (r{{(n)}^{\beta  - 3}}\right),{\rm{     \qquad 2}} < \beta  < {\rm{3}}};\\
	&{\Theta (1),{\rm{         \ \quad\qquad\qquad}}\beta  \geq 3}.
	\end{aligned}
	\end{array}} \right.
\end{equation}

$ \mathbf{Lemma\ 8.} $ When the degree of the source user is greater than $ q_{0} $, i.e., $ q>{q_0} $, the average number of hops $ E_{4} $ is
\begin{equation}\label{Lemma8}
E_{4} = \left\{ {\begin{array}{*{20}{c}}
	\begin{aligned}
	&{\Theta\left (r{{(n)}^{ - 1}}\right),{\rm{      \ \qquad 0}} \le \beta  \le {\rm{2}}};\\
	&{\Theta\left (r{{(n)}^{\beta  - 3}}\right),{\rm{    \qquad 2}} < \beta  < {\rm{3}}};\\
	&{\Theta (1),{\rm{        \ \ \ \qquad\qquad }}\beta \geq 3}.
	\end{aligned}
	\end{array}} \right.
\end{equation}

Combine Lemma 2 with Lemma 7 and Lemma 8, the result in Theorem 2 is obtained.

\section{The Upper Bound of the Capacity With Hierarchical Social Communications}
In the last section, the case with direct social communications is taken into account. However, as we mention earlier, the social communications can actually be hierarchical with multiple social levels through the inter-connected users. In other words, the source user can communicate with one of his/her $ level$-$L\ (L=1,2,...,L_{max})$ contacts through $ L-1 $ inter-users. In this section, the relationship between the extendibility of a fractal social network and the correlation exponent $ \epsilon $ is first given in Theorem 3. Then the results achieved in the last section are extended to the case with hierarchical social communications in Theorem 4.

$\mathbf{Theorem\ 3.} $  For a fractal D2D social network with $ n $ users satisfying the conditions below: 1) the $ level $-1 contacts are selected according to the joint probability distribution $ P({k_1},{k_2}) = \frac{{k_1^{ - (\gamma  - 1)}k_2^{ - \epsilon }}}{{{M_{\gamma ,\epsilon }}}},\ (k_1>k_2) $; 2) the $ level $-1 degree of each user follows the power-law degree distribution $ P(k) = \frac{{{k^{ - \gamma }}}}{{\sum_{k = 1}^n {{k^{ - \gamma }}} }} $. According to the definition given above, the relationship between the extendibility of the fractal network and the correlation exponent $ \epsilon $ is:

1) If $ 2<\epsilon<3 $, the fractal network can expand its branches continuously, and the average $ level $-$ L\ (L=1,2,...,L_{max}) $ degree increases monotonously via expanding.
 
2) $ \epsilon=3 $ is the boundary to distinguish whether the fractal network is extensible or not. In this case, the mean of $ level $-$ L $ degree keeps invariant throughout.

3) If $ \epsilon>3 $, the fractal network will stop branching rapidly after expanding finite levels. In other words, the expectation of $ level $-$ L $ degree decreases monotonously.

\emph{Proof}: Only some key elements are provided here for better readability, and the proof of Theorem 3 in detail can be found in Appendix D. 

Let $ K^{(L)} $ denote the $ level$-$L\ (L=1,2,...L_{max}) $ degree of one user, and $ {\overline K ^{(L)}} $ is the expectation of $ K^{(L)} $.

As depicted in Fig. 2(b), assume the $ level $-1 degree of Alice is known to be $ K^{(1)} $, and the crucial variable here is the average $ level $-$ L $ degree $ \overline K^{(L)} $ of Alice.

It turns out that the average $ level $-2 degree $ \overline K^{(2)} $ is:
\[
{\overline K ^{(2)}} =\frac{1}{{\epsilon-2}} \cdot \frac{{\gamma-1}}{{\gamma-2}}.
\]

When $ L\geq 3 $, the average $ level $-$ L $ degree can be derived in a similar way. Consequently, the average $ level $-$ L $ degree can be written as:
\[ 
{\overline K ^{(L)}} = \frac{1}{{\epsilon  - 2}} \cdot {\overline K ^{(L - 1)}}.
 \]

Therefore, the final expression of $\overline K^{(L)} $ is obtained:
\begin{equation}\label{Theorem3}
{\overline K ^{(L)}} = {\left( {\frac{1}{{\epsilon  - 2}}} \right)^{L - 1}} \cdot \frac{{\gamma  - 1}}{{\gamma  - 2}},\ L = 1,2,...,{L_{\max }}.
\end{equation}         

Now from Eq. \eqref{Theorem3}, it can be seen that the conclusion in Theorem 3 holds.  $\hfill\blacksquare$  

Additionally, for the convenience of understanding, the conclusions on the extendibility features in Theorem 3 are depicted in Fig. 5. As illustrated in Fig. 5, the fractal network keeps branching when $ 2<\epsilon<3 $, thus the expectation of $ level $-$ L $ degree $ \overline K ^{(L)} $ keeps rising. Specifically, the $ level $-1 degree of $ v_{i} $ is 4 while the $ level $-2 degree increases to 8 as the dotted connections in Fig. 5(a) show. The $ \overline K ^{(L)} $ remains invariant if $ \epsilon=3 $, it can be seen from the consistent $ level $-$ L $ degree of $ v_{i} $ in Fig. 5(b), which is always 4. While the expectation of degree reduces to 0 quickly when $ \epsilon>3 $ because the factor $ \frac{1}{\epsilon-2} $ is smaller than 1 in this case. As a result, the degree of $ v_{i} $ falls from 4 to 2 after the expanding of one level in Fig. 5(c).
\begin{figure}[htbp]
	\centering
	\includegraphics[scale=0.3]{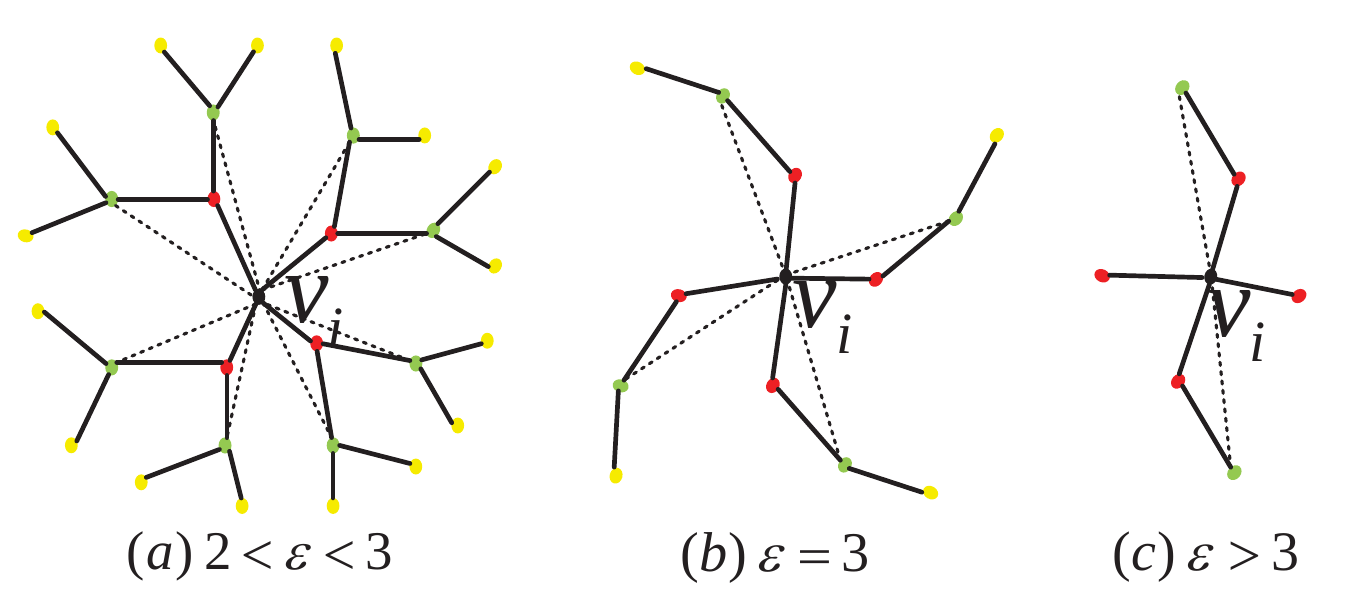}
	\caption{The extendibility of a fractal network under different values of correlation exponent $ \epsilon $.}
\end{figure}

$\mathbf{Theorem\ 4.} $ For a fractal D2D social network with $ n $ users satisfying the conditions below: 1) the $ level $-1 contacts are selected according to the joint probability distribution $ P({k_1},{k_2}) = \frac{{k_1^{ - (\gamma  - 1)}k_2^{ - \epsilon }}}{{{M_{\gamma ,\epsilon }}}},\ (k_1>k_2) $; 2) the $ level $-1 degree of each user follows the power-law degree distribution $ P(k) = \frac{{{k^{ - \gamma }}}}{{\sum_{k = 1}^n {{k^{ - \gamma }}} }} $. Then the maximum capacity $ \lambda^{(H)}_{max} $ of the fractal D2D social network with hierarchical communications is
\begin{equation}\label{Theorem4}
{\lambda ^{(H)}_{max}} = \left\{ {\begin{array}{*{20}{c}}
	\begin{aligned}
	&{\Theta \left( {\lambda_{max}  \cdot \frac{1}{{\log n}}} \right),\qquad{\rm{   2 < }}\epsilon {\rm{ < 3}}};\\
	&{\Theta (\lambda_{max}  \cdot {n^{ - 1}}),\qquad\qquad \epsilon  = 3}.
	\end{aligned}
	\end{array}} \right.
\end{equation}

\noindent where $ \lambda_{max} $ refers to the maximum capacity of fractal D2D social networks with direct communications as defined in Theorem 1 or Theorem 2 in Section III. Note that Theorem 4 holds for both uniform and power-law destination selection cases.

\emph{Proof}: Taking the social communications of all levels in a fractal D2D social network into consideration, the average number of hops $ E^{(H)}[X] $ to the destination user of arbitrary hierarchical level is
\begin{equation}\label{Theorem4.1}
\begin{array}{l}
{E^{(H)}}[X]= {E^{(1)}}[X] \cdot {R^{(1)}}+ ... + {E^{({L_{\max }})}}[X] \cdot {R^{({L_{\max }})}},
\end{array}
\end{equation}
\noindent where $ E^{(L)}[X]\ (L=1,2,...,L_{max}) $ refers to the average number of hops in the case of $ level $-$ L $ social communications. Specifically, $ E^{(1)}[X] $ is the same as the aforementioned $ E[X] $. The approximate relationship between $ E^{(L)}[X] $ and $ E^{(1)}[X] $ generally holds: $ E^{(L)}[X] \approx L \cdot E^{(1)}[X] $. In addition, $ R^{(L)}\ (L=1,2,...,L_{max}) $ stands for the ratio between the numbers of contact pairs of $ level $-$ L $ and that of all levels.

Define $ {m^{(L)}}\ (L=1,2,...,L_{max}) $ to be the number of edges in the $ level$-$L $ graph, which is formed by connecting all $ level$-$ L $ contact pairs virtually as illustrated in Fig. 3. From now on $ R^{(L)} $ is derived.

The average degree $ \overline K ^{(L)} $ and the number of edges $ m^{(L)} $ meet the equation below \cite{Mahadevan2006Systematic}: 
\[{m^{(L)}} = {\overline K ^{(L)}} \cdot n/2.\]

The total number of contact pairs of all levels is $ \left( {\begin{array}{*{20}{c}}
	n\\
	2
	\end{array}} \right) $, then the proportion of $ level $-$ L $ contact pairs is:
\[{R^{(L)}} = {m^{(L)}}/\left( {\begin{array}{*{20}{c}}
	n\\
	2
	\end{array}}\right) = \frac{{{{\overline K }^{(L)}}}}{{n - 1}}.\]

Due to the possible existence of loops in fractal D2D social networks, some contact pairs may be counted repeatedly. For instance, $ hub_{A} $ and $ hub_{B} $ in Fig. 1 can be seen as a $ level $-1 contact pair or a $ level $-4 contact pair. As a result, we define a maximum level $ L_{max} $ to imply that the contact pairs with levels larger than $ L_{max} $ have been counted before. 

Now the maximum capacity under the condition of 2$ <\epsilon< $3  and  $ \epsilon=3 $ is discussed, respectively. Define $ \alpha= \frac{1}{{\varepsilon- 2}} $ for the simplicity of the expressions.

$ \mathbf{Case\ 1} $: $ 2<\epsilon<3 $, i.e., $ \alpha > 1 $

As we know, the summation of $ R^{(L)} $ should equal to one:
\begin{equation}\label{Theorem4.2}
\begin{aligned}
&\sum\limits_{l = 1}^{{L_{\max }}} {{R^{(l)}}}  = \frac{{\gamma  - 1}}{{(\gamma  - 2)(n - 1)}}\left( {1 + \alpha + ... + {\alpha ^{{L_{\max }} - 1}}} \right)\\
&\qquad\qquad= \frac{{(\gamma  - 1)}}{{(\gamma  - 2)(\alpha  - 1)(n - 1)}} \cdot \left( {{\alpha ^{{L_{\max }}}} - 1} \right)\\
&\qquad\qquad= {\rm{1}}.
\end{aligned}
\end{equation}

So the maximum level can be calculated as:
\begin{equation}\label{Theorem4.3}
\begin{aligned}
{L_{\max }} = \frac{{\log \left[ {\frac{{(\gamma  - 2)(\alpha  - 1)(n - 1)}}{{(\gamma  - 1)}} + 1} \right]}}{{\log (\alpha )}} = \Theta \left( {\log n} \right).
\end{aligned}
\end{equation}

According to Eq. \eqref{Theorem4.1}, the average number of hops $ E^{(H)} $ in this case is:
\[\begin{aligned}
&{E^{(H)}}[X] = \sum\limits_{l = 1}^{{L_{\max }}} {{E^{(l)}}[X] \cdot {R^{(l)}}} \\
&\qquad\quad= \sum\limits_{l = 1}^{{L_{\max }}} {{E^{(1)}}[X] \cdot l \cdot \frac{{\gamma  - 1}}{{(\gamma  - 2)(n - 1)}}}  \cdot {\alpha ^{l - 1}}\\
&\qquad\quad= {E^{(1)}}[X] \cdot \frac{{\gamma  - 1}}{{(\gamma  - 2)(n - 1)}}\sum\limits_{l = 1}^{{L_{\max }}} {l \cdot } {\alpha ^{l - 1}}\\
&\qquad\quad= {E^{(1)}}[X] \cdot \frac{{\gamma  - 1}}{{(\gamma  - 2)(n - 1)}} \cdot S,
\end{aligned}\]

\noindent where $ S = \sum\limits_{l = 1}^{{L_{\max }}} {l \cdot {\alpha ^{l - 1}}} $, and it can be given analytically according to Eq. \eqref{Theorem4.2}:
\[
\begin{aligned}
&S = \frac{1}{\alpha-1}\cdot {L_{\max }} \cdot \left( {\frac{{(\gamma  - 2)(\alpha  - 1)(n - 1)}}{{\gamma  - 1}} + 1} \right) \\
&\qquad -\frac{1}{{\alpha  - 1}} \cdot \frac{{(\gamma  - 2)(n - 1)}}{{\gamma  - 1}}.
\end{aligned}
\]

According to the order of $ L_{max} $ in Eq. \eqref{Theorem4.3},
\[ 
\frac{{\gamma  - 1}}{{(\gamma  - 2)(n - 1)}} \cdot S= \Theta (\log n).
 \]

Then we can have:
\begin{equation}\label{Theorem4.4}
{E^{(H)}}[X] = \Theta \left( {{E^{(1)}}[X] \cdot \log n} \right).
\end{equation}

Combining Eq. \eqref{Theorem4.4} with Lemma 2, the maximum capacity with hierarchical social communications in the case $ 2<\epsilon<3 $ is obtained:
\begin{equation}\label{Theorem4.5}
{\lambda ^{(H)}_{max}} = \Theta \left( {\lambda_{max} \cdot \frac{1}{{\log n}}} \right).
\end{equation}

$ \mathbf{Case\ 2} $: $ \epsilon=3 $, i.e., $\alpha= \frac{1}{{\epsilon  - 2}} = 1. $ 

In this case,
\[{R^{(L)}} = \frac{{\gamma  - 1}}{{(\gamma  - 2)(n - 1)}} = {R^{(1)}}.\]

So the maximum level is calculated as:
\[{L_{\max }} = \frac{1}{{{R^{(1)}}}} = \frac{{(\gamma  - 2)(n - 1)}}{{\gamma  - 1}}=\Theta(n).\]

According to Eq. \eqref{Theorem4.1}, the average number of hops $ E^{(H)} $ is:
\[\begin{aligned}
&{E^{(H)}}[X] = \sum\limits_{l = 1}^{{L_{\max }}} {{E^{(l)}}[X] \cdot {R^{(l)}}} \\
&\qquad\qquad= {E^{(1)}}[X] \cdot {R^{(1)}} \cdot \sum\limits_{l = 1}^{{L_{\max }}} l \\
&\qquad\qquad= {E^{(1)}}[X] \cdot \frac{{(\gamma  - 2)n + 1}}{{2(\gamma  - 1)}}.
\end{aligned}\]

So the average number of hops is obtained:
\begin{equation}\label{Theorem4.6}
{E^{(H)}}[X]{\rm{ }} = \Theta \left( {{E^{(1)}}[X] \cdot n} \right).
\end{equation}

In other words, the maximum capacity in this case is:
\begin{equation}\label{Theorem4.7}
{\lambda ^{(H)}_{max}} = \Theta (\lambda_{max}  \cdot {n^{ - 1}}).
\end{equation}

To sum up Eq. \eqref{Theorem4.5} and Eq. \eqref{Theorem4.7}, the maximum capacity $ \lambda^{(H)}_{max} $ of fractal D2D social networks with hierarchical communications in Theorem 4 is achieved. $\hfill\blacksquare$

\section{Numerical Simulations And Discussions}
In this section, the results in Theorem 1, Theorem 2 and Theorem 4 are illustrated in an intuitive manner.

\begin{figure}[htbp]
		\centering
		\includegraphics[scale=0.25]{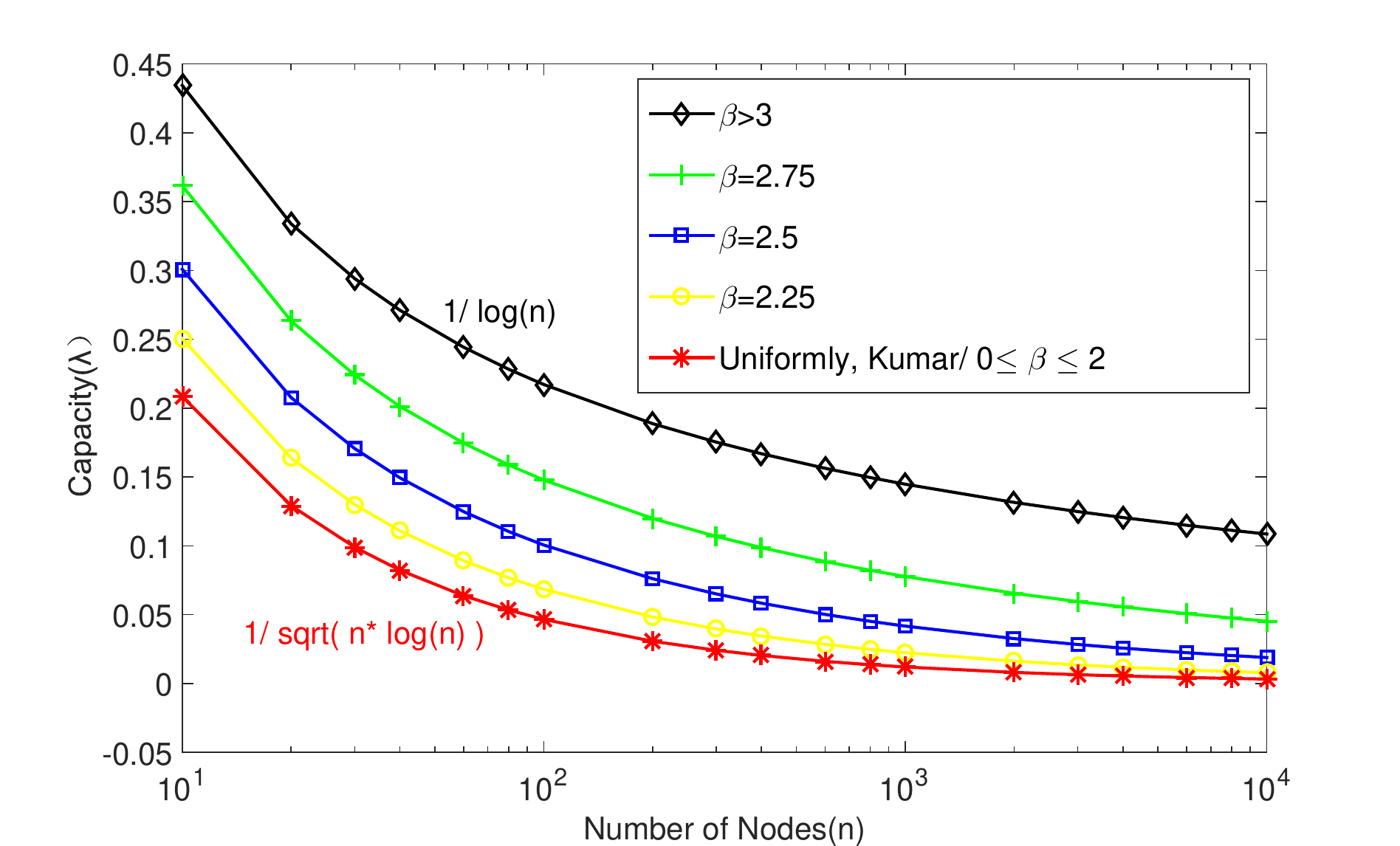}
		\caption{The maximum capacity of fractal D2D social networks with direct social communications in Theorem 1 and Theorem 2 under different values of $ \beta $.}
	\end{figure}

The direct social communications scenario is illustrated in Fig. 6, when $ \beta $ varies within $ [0,2] $, the average number of hops does not decrease distinctly compared with the uniform destination selection case. When $ \beta $ changes between $ (2,3) $, the source user prefers to communicate with closer direct social contacts, which leads to the exponentially growth of the maximum throughput. After $ \beta $ rises to 3, only $ \Theta (1) $ average hops are taken for each social transmission, which raises the maximum capacity with direct communications up to $ \Theta\left (\frac{1}{{\log n}}\right) $ finally. The results are also consistent with our findings in Theorem 2.  

Next, the effect of the correlation exponent $ \epsilon $ on the achievable capacity is discussed. According to Theorem 4, corresponding to different values of the correlation exponent $ \epsilon $, the hierarchical social communications reduce the achievable capacity, which implies the improved security level at the cost of the capacity attenuation of fractal D2D social communications. The reduction proportion is $ \frac{1}{{\log n}} $ if $ 2<\epsilon<3 $ and $ \frac{1}{n} $ if $ \epsilon=3 $, compared to that with direct social communications. The effects of reduction under two different destination selection means (i.e., uniform and power-law) are provided in Fig. 7 and Fig. 8, respectively. Specifically, it shows that the order of the maximum capacity is reduced by $ \frac{1}{log(n)} $ when the correlation exponent $ \epsilon $ is within the range $ (2,3) $, while the reduction factor is $ \frac{1}{n} $ when the correlation exponent $ \epsilon=3 $.
\begin{figure}[t]
	\centering
	\includegraphics[scale=0.23]{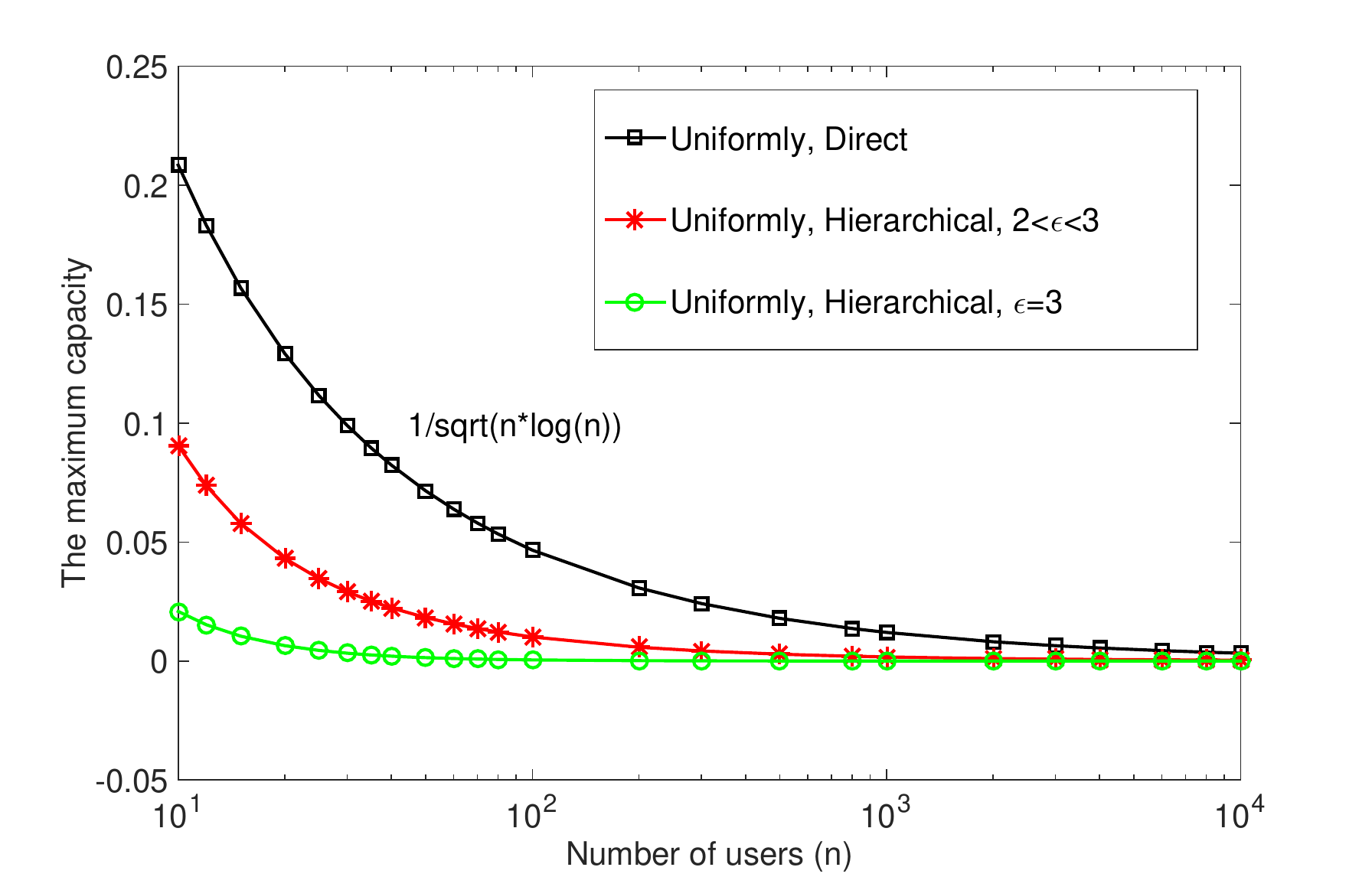}
	\caption{The comparison between the maximum capacity of fractal D2D social networks with hierarchical and direct communications under the case of uniformly distributed destinations.}
\end{figure}

\begin{figure}[t]
	\centering
	\includegraphics[scale=0.25]{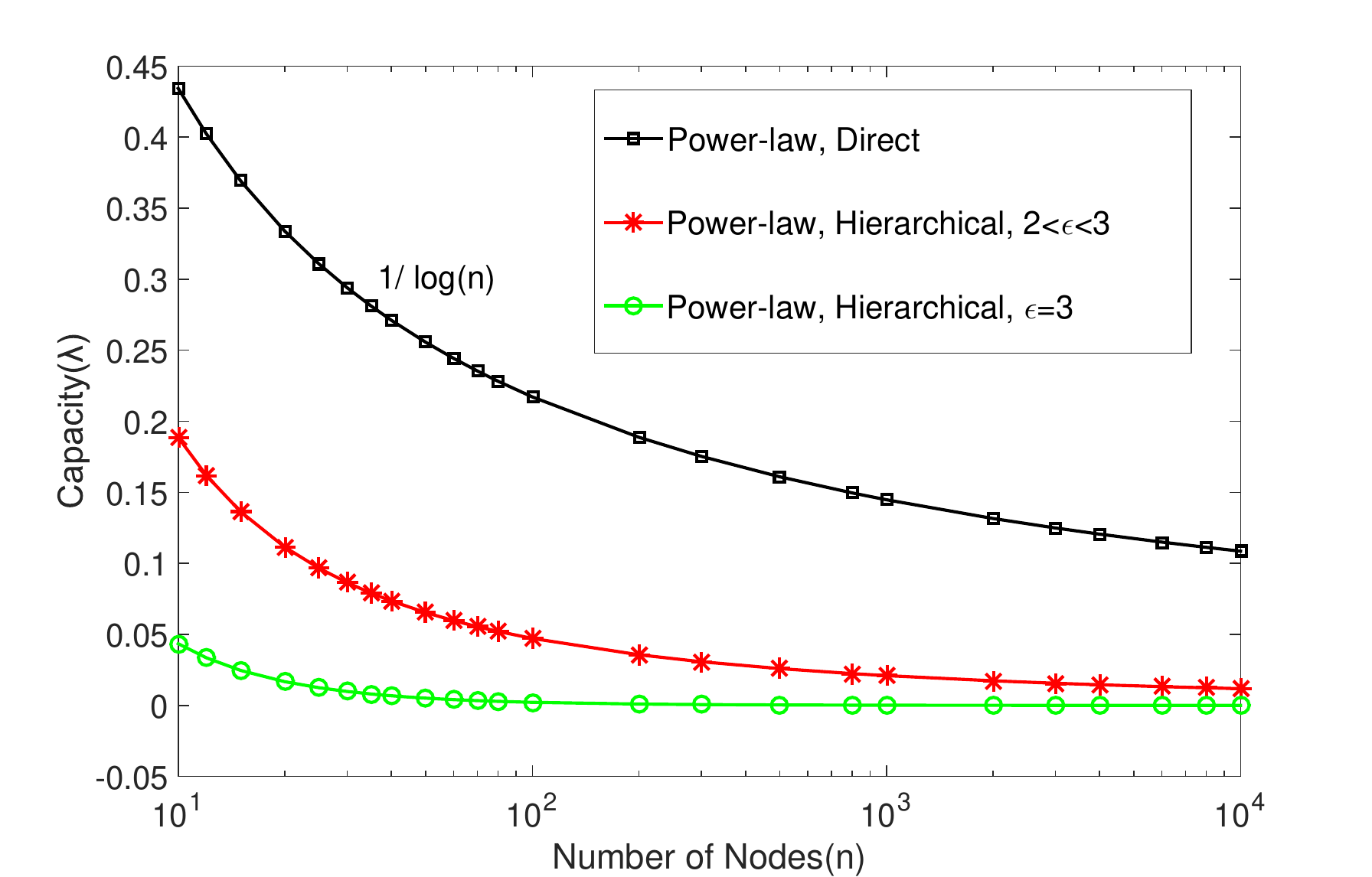}
	\caption{The comparison between the maximum capacity of fractal D2D social networks with hierarchical and direct communications under the case of power-law distributed destinations.}
\end{figure}

The reduction effect can be seen as a trade-off between the security level and achievable capacity of fractal D2D social networks, and the attenuation on the maximum capacity can be intuitively explained by the topological extension feature of fractal social networks. As is mentioned in Section II, the logarithm of the number of boxes $ N_{B}(l_{B}) $ is linearly dependent on the length scale $ l_{B} $, which indicates the possible existence of box extension with a very large size. In other words, the fractal topology stretches out the path between some social transmission pairs, which leads to the increase of the average number of hops in the scenario with hierarchical social  communications, and results in the reduction on the maximum achievable capacity naturally.

\section{Conclusion and Future Works}
In this paper, the maximum capacity of fractal D2D social networks with both direct and hierarchical communications is studied. 

Under the condition of direct social communications, it has been proved that if the source user communicates with one of his/her direct contacts randomly, the maximum capacity in Theorem 1 corresponds to the classical result $ \Theta\left(\frac{1}{\sqrt{n\log n}}\right) $ achieved by Kumar \cite{Gupta2000The}. On the other hand, if the two users with distance $ d $ communicate with each other according to the probability in proportion to $ d^{-\beta} $, the maximum capacity is
\[ {\lambda _{\max }}= \left\{ {\begin{array}{*{20}{c}}
			\begin{aligned}
			&{\Theta\left (\frac{1}{{\sqrt {n \cdot \log n} }}\right),{\rm{     \qquad\qquad\qquad 0}} \le \beta  \le {\rm{2}}};\\
			&{\Theta\left (\frac{1}{{\sqrt {{n^{3 - \beta }} \cdot \log {n^{\beta  - 1}}} }}\right),{\rm{    \qquad\quad 2}} < \beta < {\rm{3}}};\\
			&{\Theta\left (\frac{1}{{\log n}}\right),{\rm{        \ \qquad\qquad\qquad\qquad }}\beta \geq 3}.
			\end{aligned}
			\end{array}} \right. \]

While taking social communications of all levels into account, for both uniform and power-law destination selection cases, it is discovered that the hierarchical social communications further decrease the respective maximum capacity in a proportion related to the number of users $ n $, and the corresponding reduction factor varies by different values of the correlation exponent $ \epsilon $ of the fractal D2D social networks:
\[ 
{\lambda ^{(H)}_{max}} = \left\{ {\begin{array}{*{20}{c}}
	\begin{aligned}
	&{\Theta \left( {\lambda_{max}  \cdot \frac{1}{{\log n}}} \right),\qquad{\rm{   2 < }}\epsilon {\rm{ < 3}}};\\
	&{\Theta (\lambda_{max}  \cdot {n^{ - 1}}),\qquad\qquad \epsilon  = 3}.
	\end{aligned}
	\end{array}} \right.
 \]

Surely, there are still some issues remain to be solved in the future studies. For instance, why the condition $ \epsilon=3 $ is the boundary to determine whether or not the fractal network is extensible. Moreover, why is there a leap in the reduction coefficient of hierarchical social communications when $ \epsilon=3 $. We leave all these open issues in the future works.

\begin{appendix}	
\subsection{Proof of Lemma 3}
Let $ P(k=q) $ denote the probability that the degree of the source user is $ q $, while $ E[X|{\rm{source  \ }}{v_i},\ k = q] $ is the average number of hops under the condition that the source user $ v_{i} $ has $ q $ contacts, then $ E[X] $ can be written as
\begin{equation}\label{Lemma3.2}
E[X] = \sum\limits_{q=1}^{n} P(k=q)\cdot E[X|{\rm{source\ }}{v_i},\ k = q].
\end{equation}

Let $ P(X=x) $ denote the probability of $ x $ hops ranging from 1 to $ 1/r(n) $. The event $ X=x $ is true if and only if $ v_{k} $  locates in the red squares $ s_{l}\ (l=1,2,\ldots 4x) $ in Fig. 2(c) and is selected as the destination user $ v_{t} $. Therefore, $ E[X|{\rm{source \ }}{v_i},\ k = q] $ can be expanded as
\begin{equation}\label{Lemma3.3}
\begin{aligned}
&E[X|{\rm{source\ }}{v_i},\ k = q] = \sum\limits_{x=1}^{\frac{1}{r(n)}} {x \cdot } P(X = x)\\
&\qquad\qquad\qquad=\sum\limits_{x=1}^{\frac{1}{r(n)}} {x \cdot } \sum\limits_{l = 1}^{4x} {\sum\limits_{{v_k} \in {s_l}} {P({v_t} = {v_k})} }.
\end{aligned}
\end{equation}

$ \bf{C} $ is the set of all contacts of the source user. $ v_{t}=v_{k} $ implies that $ v_{k} $ is chosen as the destination user $ v_{t} $ after being selected as a contact. In other words,
\begin{equation}\label{Lemma3.4}
P(v_{t}=v_{k})=P(v_k\in\mathbf{C})\cdot P(v_t=v_k|v_k\in\mathbf{C}).
\end{equation}

Now we have the average number of hops in Eq. \eqref{Lemma3.1} by the integration of Eq. \eqref{Lemma3.2} - Eq. \eqref{Lemma3.4}:
\begin{equation}\label{Lemma3.1}
\begin{aligned}
&E[X]= \sum\limits_{q=1}^{n} P(k=q)\cdot\\
&\qquad\sum\limits_{x=1}^{\frac{1}{r(n)}} {x \cdot } \sum\limits_{l = 1}^{4x} {\sum\limits_{{v_k} \in {s_l}} {P(v_k\in\mathbf{C})\cdot P(v_t=v_k|v_k\in\mathbf{C})} }.
\end{aligned}
\end{equation}

The set $ \{v_{i_{1}},v_{i_{2}},\ldots,v_{i_{q}}\} $ contains q contacts of the source user. Taking all possible combinations into consideration, the probability that the source user has $ q $ contacts is
\[  \begin{aligned}
&P(|\mathbf{C}|=q)=\sum\nolimits_{1\leq i_{1}\leq i_{2}\leq\ldots i_{q}\leq N}P(\mathbf{C}=\{v_{i_{1}},v_{i_{2}},\ldots,v_{i_{q}}\})\\
&\qquad\qquad =\sum\nolimits_{1\leq i_{1}\leq i_{2}\leq\ldots i_{q}\leq N} \frac{(q^{-(\gamma-1)})^{q}\cdot q^{-\epsilon}_{i_{1}}q^{-\epsilon}_{i_{2}}\cdots q^{-\epsilon}_{i_{q}}}{(M_{\gamma,\epsilon})^{q}},	
\end{aligned} \]

\noindent where $ N $ is the number of users whose degree is less than $ q $ and the source user selects contacts only among these users. $ N $ grows as fast as $ n $ because
\[ 
N=n\cdot \frac{\sum^{q-1}_{b=1}b^{-\gamma}}{\sum^{n}_{b=1}b^{-\gamma}}=\Theta(n).
\]

The probability that $ \mathbf{C} $ consists of $ q $ particular users is
\[P({\mathbf{C}} = {\rm{\{ }}{v_{{i_1}}},{v_{{i_2}}}, \cdot  \cdot  \cdot, {v_{{i_q}}}\} {\rm{)}} = \frac{{q_{{i_1}}^{ - \epsilon }q_{{i_2}}^{ - \epsilon } \cdot  \cdot  \cdot q_{{i_q}}^{ - \epsilon }}}{{\sum_{1 \le {i_1} \le  \cdot  \cdot  \cdot  \le {i_q} \le N} {{\rm{    }}q_{{i_1}}^{ - \epsilon }q_{{i_2}}^{ - \epsilon } \cdot  \cdot  \cdot q_{{i_q}}^{ - \epsilon }} }}.
\] 

Consequently, the probability that $ v_{k} $ is chosen as a contact is given in Eq. \eqref{Lemma3.5} and simplified with the elementary symmetric polynomials in Definition 1 and 2:
\begin{equation}\label{Lemma3.5}
\begin{aligned}
&P({v_k} \in {\bf{C}}{\rm{)}} = \frac{{q_k^{ - \epsilon } \cdot \sum\limits_{1 \le {i_1} \le  \cdot  \cdot  \cdot  \le {i_{q - 1}} \le N} {{\rm{    }}q_{{i_1}}^{ - \epsilon }q_{{i_2}}^{ - \epsilon } \cdot  \cdot  \cdot q_{{i_{q - 1}}}^{ - \epsilon }} }}{{\sum\limits_{1 \le {i_1} \le  \cdot  \cdot  \cdot  \le {i_q} \le N} {{\rm{    }}q_{{i_1}}^{ - \epsilon }q_{{i_2}}^{ - \epsilon } \cdot  \cdot  \cdot q_{{i_q}}^{ - \epsilon }} }} \\
&\qquad\qquad= \frac{{q_k^{ - \epsilon } \cdot \sigma _{q - 1,N - 1}^{\overline k }(Q)}}{{{\sigma _{q,N}}(Q)}}.
\end{aligned}
\end{equation}

Then we have Lemma 3 after expanding Eq. \eqref{Lemma3.1} with Eq. \eqref{Lemma3.5}. $\hfill\blacksquare$ 
 
\subsection{Proof of Lemma 4}	
A transformation of the Lemma 1 suggests that
\begin{equation}\label{Lemma4.4}
\frac{{{\sigma _{1,N}}(Q)\cdot{\sigma _{q - 1,N}}(Q)}}{{q \cdot {\sigma _{q,N}}(Q)}} = \Theta (\frac{N}{{N - q + 1}}) = \Theta (1).
\end{equation}

Moreover, the probability that the degree of the source user is not greater than $ q_{0} $ is
\begin{equation}\label{Lemma4.5}
P(q \le {q_0}) = \sum\limits_{q = 1}^{{q_0}} {\frac{{{q^{ - \gamma }}}}{{\sum_{b = 1}^n {{b^{ - \gamma }}} }}}  = \Theta (1).
\end{equation}

Therefore, the upper bound of $ E_{1} $ according to Eq. \eqref{Lemma4.2} and Eq. \eqref{Lemma4.4} - Eq. \eqref{Lemma4.5} is
\begin{equation}\label{Lemma4.6}
\begin{aligned}
&E_{1} < \sum\limits_{x=1}^{\frac{1}{{r(n)}}} {x\sum\limits_{l = 1}^{4x} {\sum\limits_{{v_k} \in {s_l}} {\frac{{q_k^{ - \epsilon }\cdot{\sigma _{q - 1,N}}(Q)}}{q\cdot{{\sigma _{q,N}}(Q)}}} } } \\
&\quad \equiv \frac{{q_k^{ - \epsilon }}}{{\sigma _{1,N}}(Q)}\sum\limits_{x=1}^{\frac{1}{{r(n)}}} {x\sum\limits_{l = 1}^{4x} {\sum\limits_{{v_k} \in {s_l}} 1 } },
\end{aligned}
\end{equation}

\noindent where the symbol $ \equiv $ indicates the same order of magnitude on the two sides of an equation.

All the $ n $ users are distributed uniformly in the unit area, and the side length of each square is $ C_{1}r(n) $, so the summation term in Eq. \eqref{Lemma4.6} can be solved as
\begin{equation}\label{Lemma4.7}
\begin{aligned}
&\sum\limits_{x=1}^{\frac{1}{{r(n)}}} {x\sum\limits_{l = 1}^{4x} {\sum\limits_{{v_k} \in {s_l}}1 } }\equiv \sum\limits_{x=1}^{\frac{1}{{r(n)}}} {x \cdot 4x \cdot {C_{1}^2}{r^2}(n)}  \cdot n \cdot 1{\rm{ }}\\
&\equiv n \cdot r{(n)^2}\sum\limits_{x=1}^{\frac{1}{{r(n)}}} {{x^2} {\rm{ }}}
\equiv \Theta \left(n\cdot r{{(n)}^{ - 1}}\right).
\end{aligned}
\end{equation}

The $ q_k^{ - \epsilon } $ term in Eq. \eqref{Lemma4.6} can be replaced with its mean value in the upper bound for convenience:
\begin{equation}\label{Lemma4.8}
E[q_k^{ - \epsilon }] = \sum\limits_{b = 1}^{q - 1} {P(k = b) \cdot {b^{ - \epsilon }}}  = \frac{{\sum_{b = 1}^{q - 1} {{b^{ - (\gamma  + \epsilon )}}} }}{{\sum_{b = 1}^n {{b^{ - \gamma }}} }} \equiv \Theta (1).
\end{equation}

On the other hand,
\begin{equation}\label{Lemma4.9}
{\sigma _{1,N}}(Q) = \sum\limits_{j = 1}^N {q_j^{ - \epsilon } \equiv N \cdot \int_1^{q - 1} {{u^{ - \epsilon }}\frac{{{u^{ - \gamma }}}}{{\sum_{b = 1}^n {{b^{ - \gamma }}} }}} } du \equiv \Theta (n).
\end{equation}

By combining Eq. \eqref{Lemma4.6} - Eq. \eqref{Lemma4.9} together, the upper bound of $ E_{1} $ is obtained:
\begin{equation}\label{Lemma4.10}
E_{1} = {\rm O} \left(r{{(n)}^{ - 1}}\right). 
\end{equation}

Similarly, the lower bound of $ E_{1} $ is
\[\begin{aligned}
&{E_1} > \sum\limits_1^{\frac{1}{{r(n)}}} {x\sum\limits_{l = 1}^{4x} {\sum\limits_{{v_k} \in {s_l}} {\frac{{q_k^{ - \epsilon } \cdot {\sigma _{q - 1,N}}(Q) - q_k^{ - 2\epsilon } \cdot {\sigma _{q - 2,N}}(Q)}}{{q \cdot {\sigma _{q,N}}(Q)}}} } }
{\rm{   }} \\
&\quad= {\rm{upper\ bound - }}\sum\limits_1^{\frac{1}{{r(n)}}} {x\sum\limits_{l = 1}^{4x} {\sum\limits_{{v_k} \in {s_l}} {\frac{{q_k^{ - 2\epsilon } \cdot {\sigma _{q - 2,N}}(Q)}}{{q \cdot {\sigma _{q,N}}(Q)}}} } }.
\end{aligned} \]

It turns out that the second term in the lower bound is
\begin{equation}\label{Lemma4.11}
\sum\limits_{x=1}^{\frac{1}{{r(n)}}} {x\sum\limits_{l = 1}^{4x} {\sum\limits_{{v_k} \in {s_l}} {\frac{{q_k^{ - 2\epsilon }\cdot{\sigma _{q - 2,N}}(Q)}}{q\cdot{{\sigma _{q,N}}(Q)}}} } } \equiv \Theta \left(n^{-1}\cdot r{{(n)}^{ - 1}}\right).
\end{equation}

The order in Eq. \eqref{Lemma4.11} is negligible compared with the upper bound in Eq. \eqref{Lemma4.10}, so the order of $ E_{1} $ in Eq. \eqref{Lemma4.3} is solved.      $\hfill\blacksquare$

\subsection{Proof of Lemma 5}
Similar to the case $ E_{1} $, $ E_{2} $ is given as
\begin{equation}\label{Lemma5.1}
{E_2} = \sum\limits_{q = {q_0} + 1}^n {\frac{{{q^{ - \gamma }}}}{{\sum_{b = 1}^n {{b^{ - \gamma }}} }}}  \cdot \sum\limits_{x=1}^{\frac{1}{{r(n)}}} {x\sum\limits_{l = 1}^{4x} {\sum\limits_{{v_k} \in {s_l}} {\frac{{q_k^{ - \epsilon }\cdot\sigma _{q - 1,N - 1}^{\overline k }(Q)}}{q\cdot {{\sigma _{q,N}}(Q)}}} } } .
\end{equation}

Since $ N $ is large enough and the degrees of $ q $ social contacts are independent and identically distributed, the law of large numbers can work here. Let $ {{\rm{X}}_{{i_j}}} = q_{{i_j}}^{ - \epsilon },\ {{\rm{Y}}_{{i_j}}} = \log {X_{{i_j}}} $, and $ \overline Y  $ denote the mean of $ Y_{i_{j}} $, then we have
\begin{equation}\label{Lemma5.2}
\begin{aligned}
&\frac{{q_k^{ - \epsilon }\cdot\sigma _{q - 1,N - 1}^{\overline k }(Q)}}{q\cdot{{\sigma _{q,N}}(Q)}} \equiv \frac{{\sum_{1 \le {i_1} \le  \cdot  \cdot  \cdot  \le {i_q} \le N,\exists m,{i_m} = k} {\prod_{{\rm{j}} = {\rm{1}}}^{\rm{q}} {{{\rm{X}}_{{{\rm{i}}_{\rm{j}}}}}} } }}{q\cdot{\sum_{1 \le {i_1} \le  \cdot  \cdot  \cdot  \le {i_q} \le N} {\prod_{{\rm{j}} = {\rm{1}}}^{\rm{q}} {{{\rm{X}}_{{{\rm{i}}_{\rm{j}}}}}} } }}\\
&\qquad\qquad\qquad\ \equiv\frac{{\sum_{1 \le {i_1} \le  \cdot  \cdot  \cdot  \le {i_q} \le N,\exists m,{i_m} = k} {\exp (\sum_{{\rm{j}} = {\rm{1}}}^{\rm{q}} {{{\rm{Y}}_{{{\rm{i}}_{\rm{j}}}}})} } }}{q\cdot{\sum_{1 \le {i_1} \le  \cdot  \cdot  \cdot  \le {i_q} \le N} {\exp (\sum_{{\rm{j}} = {\rm{1}}}^{\rm{q}} {{{\rm{Y}}_{{{\rm{i}}_{\rm{j}}}}})} } }}\\
&\qquad\qquad\qquad\ \equiv \frac{{\sum_{1 \le {i_1} \le  \cdot  \cdot  \cdot  \le {i_q} \le N,\exists m,{i_m} = k} {{\rm{        }}\exp (q\overline {\rm{Y}} )} }}{q\cdot{\sum_{1 \le {i_1} \le  \cdot  \cdot  \cdot  \le {i_q} \le N} {{\rm{    }}\exp (q\overline {\rm{Y}} )} }}\\
&\qquad\qquad\qquad\ \equiv \frac{{\left( {\begin{array}{*{20}{c}}
			{N - 1}\\
			{q - 1}
			\end{array}} \right)}}{q\cdot{\left( {\begin{array}{*{20}{c}}
			N\\
			q
			\end{array}} \right)}} = \frac{1}{N} = \Theta (n^{-1}).
\end{aligned}
\end{equation}

Besides, the probability that the degree of the source user is greater than $ q_{0} $ is 
\begin{equation}\label{Lemma5.3}
P(q > {q_0}) = \sum\limits_{q = {q_0} + 1}^n {\frac{{{q^{ - \gamma }}}}{{\sum_{b = 1}^n {{b^{ - \gamma }}} }}}  = \Theta (1).
\end{equation}

Then Eq. \eqref{Lemma5.1} can be simplified by Eq. \eqref{Lemma5.2} - Eq. \eqref{Lemma5.3}, namely:
\begin{equation}\label{Lemma5.4}
E_{2} \equiv \sum\limits_{x=1}^{\frac{1}{{r(n)}}} {x\sum\limits_{l = 1}^{4x} {\sum\limits_{{v_k} \in {s_l}} {\frac{1}{n}} } }\equiv\Theta \left(r{{(n)}^{ - 1}}\right).
\end{equation}

Therefore, Lemma 5 is proved. $\hfill\blacksquare$

\subsection{Proof of Theorem 3}
Before giving the proof, the definition of moment generating function\cite{Simon2006Capacity} and its properties need to be introduced.

$ \mathbf{Definition\ 3.} $ For a discrete random variable X, its moment generating function is defined as:
\[{\phi _X}(t) = E[{e^{tX}}] = \sum\limits_{x = 0}^\infty  {{e^{tx}}}  \cdot P(X = x).\]

In addition, it is easy to obtain some useful properties of moment generating function:

$ \mathbf{Property\ 1:} $ $ {\phi _X}(0) = \sum\limits_{x = 0}^\infty  {P(X = x) = 1} $.

$ \mathbf{Property\ 2:} $ $ \phi {'_X}(0) = E[X] $.

$ \mathbf{Property\ 3:} $ For the discrete random variables $ X $, $ Y $ and $ Z $, if $ Z = X + Y $, and $ X $ and $ Y $ are independent of each other, then $ {\phi _Z}(t) = {\phi _X}(t) \cdot {\phi _Y}(t) $.

Now, Theorem 3 can be proved.

In the first place, some symbols are defined: $ {M_\gamma } = \sum\limits_{k = 1}^n {{k^{ - \gamma }}}  $ and $ {M_\epsilon } = \sum\limits_{k = 1}^n {{k^{ - \epsilon }}}  $ are two normalization factors; $ {\overline K ^{(L)}} $ denotes the average degree of $ level$-$L\ (L=1,2,...L_{max}) $; each social connection has two end users, and $ D $ stands for the degree of one end user when that of another end user is known. When the degree of Alice is known in Fig. 2(b), for example, the degree distribution of Bob follows the power-law $ P(D = k) = \frac{{{k^{ - \epsilon }}}}{{{M_\epsilon }}},\ k=1,2,...n $ according to $ P(k_{1},k_{2}) $; $ \overline D $ is the expectation of $ D $. 

Hereinafter the focus is the order of capacity. When $ n $ goes to positive infinity, some of their values can be calculated as below since $ \gamma $ and $ \epsilon $ are both greater than $ 2 $:
\[
\begin{aligned}
&{M_\gamma } = \sum\limits_{k = 1}^n {{k^{ - \gamma }}}  = \frac{{{k^{1 - \gamma }}}}{{\gamma  - 1}}\bigg|_n^1 \approx \frac{1}{{\gamma  - 1}},\\
&{M_\epsilon } = \sum\limits_{k = 1}^n {{k^{ - \epsilon }}}  = \frac{{{k^{1 - \epsilon }}}}{{\epsilon  - 1}}\bigg|_n^1 \approx \frac{1}{{\epsilon  - 1}},\\
&{\overline K ^{(1)}} = E[{K^{(1)}}] = \sum\limits_{k = 1}^n {k \cdot } P({K^{(1)}} = k)\approx \frac{{\gamma  - 1}}{{\gamma  - 2}},\\
&\overline D  = E[D] = \sum\limits_{k = 1}^n {k \cdot P(D = k) = } \sum\limits_{k = 1}^n {k \cdot \frac{{{k^{ - \epsilon }}}}{{{M_\epsilon }}} \approx } \frac{{\epsilon  - 1}}{{\epsilon  - 2}}.
\end{aligned}
\] 

As depicted in Fig. 2(b), the $ level $-1 degree of Alice is known to be $ K^{(1)} $, and we intend to solve the average $ level $-2 degree $ K^{(2)} $ of Alice, which can be expressed as:
\[{K^{(2)}} = ({D_1} - 1) + ({D_2} - 1) + ... + ({D_{{K^{(1)}}}} - 1),\]

\noindent where $ {D_1},{D_2},...,{D_{{K^{(1)}}}} $ are independently and identically power-law distributed as $ P(D_i= k) =P(D= k)= \frac{{{k^{ - \epsilon }}}}{{{M_\epsilon }}},\ i=1,2,...,K^{(1)},\ k=1,2,...n$. The return connection is subtracted from each of $ D_{i} $ to alleviate the impact of the loops. And $ K^{(1)} $ follows the aforementioned power-law distribution $ P(k) $, i.e., $ P({K^{(1)}} = k) = \frac{{{k^{ - \gamma }}}}{{{M_\gamma }}},\ k=1,2,...n $.

Define $ K{'^{(2)}} = \sum\limits_{i = 1}^{{K^{(1)}}} {{D_i}} $, then $ {K^{(2)}} = K{'^{(2)}} - {K^{(1)}} $.

Under the condition of $ K^{(1)} $, the moment generating function of $ K{'^{(2)}} $ is written as:
\[\begin{array}{l}
{\phi _{K{'^{(2)}}}}(t) = E[{e^{tK{'^{(2)}}}}]= \sum\limits_{k = 1}^n {P({K^{(1)}}}  = k) \cdot E[{e^{tK{'^{(2)}}}}|{K^{(1)}} = k].
\end{array}\]

According to the Property 3 of moment generating function:
\[\begin{array}{l}
E[{e^{tK{'^{(2)}}}}|{K^{(1)}} = k] = {\phi _{{D_1}}}(t) \cdot {\phi _{{D_2}}}(t) \cdot ... \cdot {\phi _{{D_k}}}(t)= {[{\phi _D}(t)]^k}.
\end{array}\]

Then $ \phi_{K{'^{(2)}}}(t) $ is:
\[{\phi _{K{'^{(2)}}}}(t) = \sum\limits_{k = 1}^n {P({K^{(1)}}}  = k) \cdot {[{\phi _D}(t)]^k}.\]

Now calculate the expectation of $ K{'^{(2)}} $ by the Property 1 and 2 of moment generating function:
\[\begin{aligned}
&E[K{'^{(2)}}] = \phi {'_{K{'^{(2)}}}}(t){|_{t = 0}}\\
&\qquad\quad= \sum\limits_{k = 1}^n {k \cdot P({K^{(1)}} = k) \cdot {{[{\phi _D}(t)]}^{k - 1}}}  \cdot \phi {'_D}(t){|_{t = 0}}\\
&\qquad\quad = \sum\limits_{k = 1}^n {k \cdot P({K^{(1)}} = k) \cdot {1^{k - 1}}}  \cdot \overline D \\
&\qquad\quad= \overline D  \cdot {\overline K ^{(1)}}.
\end{aligned}\]

Therefore, the average $ level $-2 degree $ {\overline K ^{(2)}} $ is
\[\begin{aligned}
&{\overline K ^{(2)}} = E[K{'^{(2)}} - {K^{(1)}}] = (\overline D  - 1) \cdot {\overline K ^{(1)}}
&= \frac{1}{{\epsilon-2}} \cdot \frac{{\gamma-1}}{{\gamma-2}}.
\end{aligned}\]

When $ L\geq 3 $, the average $ level $-$ L $ degree can be derived in the same way. The $ level $-$ L $ degree $ {K^{(L)}} $ can be expanded as:
\[{K^{(L)}} = ({D_1} - 1) + ({D_2} - 1) + ... + ({D_{{K^{(L - 1)}}}} - 1) = K{'^{(L)}} - {K^{(L - 1)}},\]

\noindent where the variables $ {D_1},{D_2},...,{D_{{K^{(L - 1)}}}} $ are independently and identically distributed as $ P(D=k) $, and $ K{'^{(L)}} = \sum\limits_{i = 1}^{{K^{(L - 1)}}} {{D_i}} $.

The moment generating function of $K{'^{(L)}} $ is:
\[\begin{aligned}
&{\phi _{K{'^{(L)}}}}(t) = E[{e^{tK{'^{(L)}}}}]\\
&\qquad\quad = \sum\limits_{k = 1}^n {P({K^{(L - 1)}}}  = k) \cdot E[{e^{tK{'^{(L)}}}}|{K^{(L - 1)}} = k],
\end{aligned}\]

\noindent and the expectation of $ K{'^{(L)}} $ can be solved as:
\[\begin{aligned}
&E[K{'^{(L)}}] = \phi {'_{K{'^{(L)}}}}(t){|_{t = 0}}\\
&\qquad\quad = \sum\limits_{k = 1}^n {k \cdot P({K^{(L - 1)}} = k) \cdot {{[{\phi _D}(t)]}^{k - 1}}}  \cdot \phi {'_D}(t){|_{t = 0}}\\
&\qquad\quad = \sum\limits_{k = 1}^n {k \cdot P({K^{(L - 1)}} = k) \cdot {1^{k - 1}}}  \cdot \overline D \\
&\qquad\quad = \overline D  \cdot {\overline K ^{(L - 1)}}.
\end{aligned}\]

That is to say, the average $ level $-$ L $ degree can be written as:
\[\begin{aligned}
&{\overline K ^{(L)}} = E[K{'^{(L)}} - {K^{(L - 1)}}]\\
&\qquad = (\overline D  - 1) \cdot {\overline K ^{(L - 1)}}\\
&\qquad = \frac{1}{{\epsilon  - 2}} \cdot {\overline K ^{(L - 1)}}.
\end{aligned}\]

Therefore, the final expression of $\overline K^{(L)} $ in Eq. \eqref{Theorem3} is obtained:
\[ 
{\overline K ^{(L)}} = {\left( {\frac{1}{{\epsilon  - 2}}} \right)^{L - 1}} \cdot \frac{{\gamma  - 1}}{{\gamma  - 2}},\ L = 1,2,...,{L_{\max }}.
 \]                    $\hfill\blacksquare$

\end{appendix}

\bibliographystyle{IEEEtran}
\bibliography{IEEEfull,reference}
\end{document}